\documentclass{jpsj2}
\usepackage{graphicx}
\usepackage{amsmath,amssymb}

\def\runtitle
{
Microscopic Identification of the Pairing State in Triplet Superconductor 
${\rm Sr_{2}RuO_{4}}$
}
\def\runauthor{Youichi {\sc Yanase} and Masao {\sc Ogata}}

\title{ 
Microscopic Identification of the $D$-vector in Triplet Superconductor 
${\rm Sr_{2}RuO_{4}}$
} 

\author{Youichi {\sc Yanase}\footnote{E-mail: yanase@hosi.phys.s.u-tokyo.ac.jp}and Masao {\sc Ogata}}

\inst{Department of Physics, University of Tokyo, Tokyo 113-0033}

\recdate{19 August 2002}

\abst
{
 Triplet superconductivity in ${\rm Sr}_{2}{\rm Ru}{\rm O}_{4}$ is 
investigated with main interest on its internal degree of freedom. 
 We perform a microscopic calculation to investigate how the chiral state 
$\hat{d}(k) = (k_{{\rm x}} \pm {\rm i} k_{{\rm y}}) \hat{z}$ is realized 
among the underlying six degenerate states. 
 Starting from the three band Hubbard model with spin-orbit interaction, 
we use a perturbation theory in order to calculate the pairing interaction. 
 The $p$-wave superconductivity with $T_{\rm c} \sim 1.5{\rm K}$ 
is obtained in the moderately weak coupling region. 
 It is shown that the orbital dependent superconductivity (ODS) 
robustly appears in ${\rm Sr}_{2}{\rm Ru}{\rm O}_{4}$. 
 We determine the stabilized state 
by solving the Eliashberg equations. 
 We find that the Hund coupling term as well as the spin-orbit interaction 
is necessary for the ``symmetry breaking interaction''. 
 The main result is that the chiral state is stabilized in case of 
the $p$-wave symmetry with the main $\gamma$-band, 
which is obtained in the perturbation theory. 
 When we assume the other pairing symmetry including the $f$-wave state, 
the symmetry breaking interaction gives the other $d$-vector.  
 The electronic structure constructed from the $t_{{\rm 2g}}$-orbitals is 
essential for this result. 
}

\kword
{Spin triplet superconductivity; ${\rm Sr}_{2}{\rm Ru}{\rm O}_{4}$;  
D-vector; spin-orbit coupling
}

\begin{document}
\sloppy
\maketitle

\newcommand{\eli}{$\acute{{\rm E}}$liashberg }
\renewcommand{\k}{\mbox{\boldmath$k$}}
\newcommand{\q}{\mbox{\boldmath$q$}}
\newcommand{\Q}{\mbox{\boldmath$Q$}}
\newcommand{\kk}{\mbox{\boldmath$k'$}}
\newcommand{\e}{\varepsilon}
\newcommand{\ee}{\varepsilon^{'}}
\newcommand{\s}{{\mit{\it \Sigma}}}
\newcommand{\J}{\mbox{\boldmath$J$}}
\newcommand{\vv}{\mbox{\boldmath$v$}}
\newcommand{\Jh}{J_{{\rm H}}}
\newcommand{\LL}{\mbox{\boldmath$L$}}
\renewcommand{\SS}{\mbox{\boldmath$S$}}
\newcommand{\Tc}{$T_{\rm c}$}

\section{Introduction}

 The discovery~\cite{rf:maeno} and establishment~\cite{rf:maenoPT} of spin 
triplet superconductivity in ${\rm Sr}_{2}{\rm Ru}{\rm O}_{4}$ have stimulated 
much interests. 
 ${\rm Sr}_{2}{\rm Ru}{\rm O}_{4}$ has the same crystal structure as 
high-$T_{{\rm c}}$ cuprates and possesses a quasi-two-dimensional 
nature.~\cite{rf:mackenzie,rf:maenoFermi2D} 
 In contrast to high-$T_{{\rm c}}$ cuprates,~\cite{rf:timusk,rf:yanasePG} 
this system behaves as a typical two-dimensional Fermi liquid in the normal 
state.~\cite{rf:mackenzie,rf:maenoFermi2D}  
 Three cylindrical Fermi surfaces are clearly observed by the quantum 
oscillation measurement.~\cite{rf:mackenzie} 
 These conduction band originates from the 
$t_{{\rm 2g}}$-orbitals in Ru ions.~\cite{rf:oguchi,rf:singh} 
 The $T^{2}$-law of the resistivity~\cite{rf:maenoFermi2D} 
and $3 \sim 4$ times mass enhancement~\cite{rf:mackenzie} indicate  
the importance of the electron correlation.  
 The correlation effect is also suggested 
from the Mott insulating state in the related compound 
${\rm Ca}_{2}{\rm Ru}{\rm O}_{4}$.~\cite{rf:nakatsuji}

 Theoretical interests have been focused on the unconventional 
superconductivity  expected from the correlation effect and 
the low-dimensionality. 
 The possibility of spin triplet superconductivity was first pointed out 
by Rice and Sigrist,~\cite{rf:rice} which was succeedingly supported by 
many experimental results. 
 For example, NMR and NQR studies have shown no Hebel-Slichter 
peak,~\cite{rf:ishidaH} 
 the transition temperature ($T_{{\rm c}}$) is suppressed by non-magnetic 
impurities,~\cite{rf:mackenzieimp} and the time-reversal symmetry breaking 
is observed by $\mu{\rm SR}$.~\cite{rf:luke} 
 The most significant evidence for spin triplet superconductivity is obtained  
by the NMR Knight shift~\cite{rf:ishidaK} and by the neutron scattering 
measurements,~\cite{rf:duffy} which show that the spin susceptibility 
does not change through $T_{{\rm c}}$. 
 The singlet pairing is denied by this fact.

 The triplet superconductivity has already been observed in heavy Fermion 
compounds, such as ${\rm UPt}_{3}$~\cite{rf:stewart}. Probably, the recently 
discovered compounds ${\rm UGe}_{2}$~\cite{rf:saxena} and 
URhGe~\cite{rf:Aoki} are also the triplet superconductors. 
 However, the microscopic investigation in these materials is generally 
difficult because of their complicated electronic structure. 
 Actually, any microscopic theory acceptable for the superconductivity in 
heavy Fermion systems has not been constructed. 
 On the other hand, the electronic structure of 
${\rm Sr}_{2}{\rm Ru}{\rm O}_{4}$ is relatively simple, that is, 
it has two-dimensional Fermi surfaces, 
a few degenerate orbitals, weak spin-orbit coupling and 
weak electron correlation. 
 Therefore, ${\rm Sr}_{2}{\rm Ru}{\rm O}_{4}$ is the most precious and 
favorable compound for the microscopic investigation on the triplet 
superconductivity. 
 Such studies on ${\rm Sr}_{2}{\rm Ru}{\rm O}_{4}$ will give  
various information also for the unconventional superconductivity  
in other materials.

 There have been several proposals for the microscopic calculation for the 
triplet superconductivity in ${\rm Sr}_{2}{\rm Ru}{\rm O}_{4}$, 
including the random phase approximation (RPA)~\cite{rf:takimoto,rf:ogata} 
and perturbative method~\cite{rf:nomura,rf:nomuramulti}. 
 In this paper, we use the perturbative method which will be justified in the 
moderately weak coupling region and be complementary to the RPA theories. 
 If we use the spin isotropic (SU(2) symmetric) model, the possible triplet 
Cooper pairings are degenerate. In this paper, we take account of 
the spin-orbit interaction in the three band Hubbard-type model to lift the 
degeneracy, and show that the chiral state, which is consistent with 
experiments, is stabilized under the reasonable parameters (\S3). 
 In this case, the $\gamma$-band is mainly superconducting, 
and the $k$-dependence of the order parameter is classified 
into the $p$-wave symmetry (\S2). 
 It is shown that this pairing symmetry is essential to stabilize 
the $d$-vector along the {\it z}-axis (\S4).

 The results obtained in this paper are related to the several interesting 
issues for ${\rm Sr}_{2}{\rm Ru}{\rm O}_{4}$; 
(a) direction of the $d$-vector (b) pairing symmetry in $\k$-space 
(c) the power-law behaviors in various quantities,~\cite{rf:nishizaki,
rf:bonalde,rf:ishidanode,rf:tanatar1,rf:lupien} and (d) mechanism of 
the triplet superconductivity. 
 We briefly review these issues and the theoretical proposals for them. 

(a) $d$-vector. 

 In the triplet superconductivity, the internal degree of freedom is an 
attractive character, {\it i.e.}, the order parameter has three components 
in the spin space and is described by the $d$-vector as $\Delta(k)= {\rm i} 
\hat{d}(k) \hat{\sigma} \sigma_{y}$.~\cite{rf:leggett,rf:sureview} 
 However, 
the identification of the $d$-vector using the microscopic Hamiltonian 
is usually difficult, and actually not performed. 
 Instead, it has been discussed phenomenologically~\cite{rf:sureview,
rf:leggett,rf:ohmi,rf:sauls,rf:sigristreview}.

 In case of ${\rm Sr}_{2}{\rm RuO}_{4}$, there exists a six-fold degeneracy 
together with the momentum component when the spin-orbit interaction 
is neglected.~\cite{rf:sigristreview} 
 The experimental results have concluded that the chiral state 
$\hat{d}(k) = (k_{{\rm x}} \pm {\rm i} k_{{\rm y}}) \hat{z}$ is realized among 
the six-fold degeneracy, in which the time-reversal symmetry is 
broken.~\cite{rf:luke,rf:ishidaK}~\footnote{
Strictly speaking, this result relies on the $\mu$SR experiment 
because the NMR Knight shift is measured under the parallel magnetic field. 
 The NMR under the perpendicular magnetic field is desired for a stronger 
evidence. 
}

 In order to explain why the $d$-vector is parallel to $\hat{z}$-axis, 
phenomenological calculations have been proposed by assuming the momentum 
dependence of the quasi-particle's interaction~\cite{rf:ngls} 
or the anisotropy of the spin fluctuation.~\cite{rf:kuwabara,rf:satoh,
rf:kuroki} 
 In this paper, we carry out a microscopic calculation starting from the 
Hubbard-type Hamiltonian to identify the $d$-vector. 
 The microscopic origin and the general properties of the 
``symmetry breaking interaction'' are investigated. 
 We find that the dominant contribution comes from the terms 
which have been overlooked in the phenomenological theories.

(b) pairing symmetry. 

 The issue on the pairing symmetry in $\k$-space is closely related to (c) the 
power-law behaviors.~\cite{rf:nishizaki,rf:bonalde,rf:ishidanode,rf:tanatar1,
rf:lupien}   
 Because a finite excitation gap is expected in a pure $p$-wave chiral state, 
the understanding of the gap-less behaviors is one of the main issues 
at the present stage. 
 Actually, many theoretical studies have been dedicated to 
the resolution of this inconsistency.

 In order to compare various proposals, we explicitly write the momentum 
dependence of the order parameter. 
 The chiral state is described as 
$\hat{d}(k) = (\phi_{{\rm x}}(k) \pm {\rm i} \phi_{{\rm y}}(k)) \hat{z}$ 
where $\phi_{{\rm x}}(k)$ ($\phi_{{\rm y}}(k)$) is a wave function 
having odd symmetry with respect to the reflection  
$k_{{\rm x}} \rightarrow -k_{{\rm x}}$ 
($k_{{\rm y}} \rightarrow -k_{{\rm y}}$). 
 For example, the tiny gap model has been proposed by assuming the 
$p$-wave function $\phi_{{\rm x}}(k) = \sin k_{{\rm x}}$.~\cite{rf:miyake} 
 Another candidate is the 2D $f$-wave symmetry; the $f_{x^{2}-y^{2}}$-wave 
state~\cite{rf:dahm} ($\phi_{{\rm x}}(k)=
\sin k_{{\rm x}}(\cos k_{{\rm x}}-\cos k_{{\rm y}})$) 
and $f_{xy}$-wave state~\cite{rf:graf}
($\phi_{{\rm x}}(k) = \sin k_{{\rm x}} \sin k_{{\rm y}}^{2}$) 
have been proposed. 
 In relation with the thermal conductivity measurements,~\cite{rf:izawa,
rf:tanatar} the 3D $f$-wave state with a horizontal line node 
has been proposed.~\cite{rf:hasegawa,rf:won,rf:zhitomirski}

 Although these proposals have succeeded in explaining the gap-less behaviors 
at least phenomenologically, a consensus remains to be obtained. 
 In this paper, we compare the above gap functions including the orbital 
dependence by investigating whether chiral state is favored or not. 
 It will be shown that the symmetry-breaking interaction given 
by the perturbation theory favors the chiral state {\it only when} 
the $\gamma$-band is mainly superconducting with $p$-wave symmetry. 
 For example, another $d$-vector 
$\hat{d}(k) = \phi_{x}(k) \hat{y} \pm \phi_{y}(k) \hat{x}$ is obtained 
in case of the $f_{x^{2}-y^{2}}$-wave symmetry.

 Recently, it was proposed that the power-law behaviors can be explained 
by the multi-band effect in a particular parameter set.~\cite{rf:nomuratherm}  
 Then, three ideas play an essential role: 
(i) anisotropic gap structure in $\gamma$-band~\cite{rf:miyake}, 
(ii) orbital dependent superconductivity (ODS)~\cite{rf:agterberg} and 
most importantly 
(iii) node-like structure in $\beta$-band.~\cite{rf:kuroki,rf:nomuratherm} 
 Here, we use ``ODS'' as a situation where the amplitude of 
the order parameter strongly depends on the orbital. 
 This proposal is consistent with the above argument on the $d$-vector. 
 Although we do not discuss this possibility, we find that the ODS naturally 
appears in ${\rm Sr}_{2}{\rm RuO}_{4}$.

(d) mechanism of the triplet superconductivity.

 The ferromagnetic spin fluctuations were first 
speculated~\cite{rf:rice,rf:sigrist,rf:mazinferro,rf:monthoux} in analogy with 
the superfluid $^{3}{\rm He}$.~\cite{rf:anderson,rf:leggett} 
 This proposal, however, has been denied by the inelastic neutron scattering 
showing no sizable ferromagnetic spin fluctuations.~\cite{rf:sidis} 
 On the contrary, the neutron scattering detected 
the incommensurate spin fluctuations around 
$\q \sim (\frac{2}{3}\pi, \frac{2}{3}\pi)$, which were predicted in the 
band calculation~\cite{rf:mazinincom} and in the 
multi-band RPA.~\cite{rf:nomuraHF} 
 The incommensurate fluctuations have an anisotropic 
nature~\cite{rf:ishidaanisotropy}~\footnote{A query on this experimental 
result has been raised by the neutron scattering experiment.~\cite{rf:servant}}
$\chi_{{\rm z}}(\q) > \chi_{\pm}(\q)$ 
which is derived from the spin-orbit interaction.~\cite{rf:ngmagne} 
 It has been proposed that this anisotropic spin fluctuations can produce the 
$p$-wave superconductivity with $d$-vector parallel to 
{\it z}-axis.~\cite{rf:kuwabara,rf:kuroki,rf:satoh} 
 Since the incommensurate fluctuations are derived from the nesting of the 
$\alpha$ and $\beta$ Fermi surface, this mechanism is favorable in the 
$\alpha$- and $\beta$-bands.~\cite{rf:ogata} 
 However, the more detailed calculation will be necessary since the 
$\gamma$-band has the largest density of states (DOS).

 Another proposal is the orbital fluctuations with the same 
wave vector $\q \sim (\frac{2}{3}\pi, \frac{2}{3}\pi)$.  
 Takimoto has shown that the $f$-wave superconductivity is induced mainly on 
the $\alpha$- and $\beta$-band.~\cite{rf:takimoto} 
 This mechanism requires that the inter-orbit repulsion is larger than the 
intra-orbit one.  
 In general, this condition is difficult to be satisfied.

 Nomura and Yamada have proposed an another approach based on the 
perturbation theory.~\cite{rf:nomura,rf:nomuramulti} 
 This approach is based on the general concept for the electronic 
mechanism.~\cite{rf:kohn} 
 The calculation for the single band model within the third order has given 
the $p$-wave superconductivity.~\cite{rf:nomura}. 
 It has been shown that the consistent results are obtained 
for the three band model where the $\gamma$-band is mainly 
superconducting.~\cite{rf:nomuramulti}

 In this paper, we similarly use the perturbation theory and 
show that the $p$-wave superconductivity with $T_{{\rm c}} \sim 1.5 {\rm K}$ 
is obtained in the moderately weak coupling region (\S2.4). 
 In this region, the third order term is sufficiently smaller than 
the second order term, but enhances the triplet superconductivity 
in cooperation with the second order term. 
 The perturbative treatment for the ``symmetry breaking interaction'' 
consistently gives the chiral state. 
 Thus, the triplet superconductivity with appropriate 
$d$-vector is obtained in the perturbative region when the characteristic 
band structure is correctly taken into account.

 Hereafter, we restrict the discussion to the unitary state 
in the 2D system. 
 The $k_{{\rm z}}$-dependence of the order parameter, and therefore 
the 3D f-wave state are not considered. 
 Generally speaking, the 3D f-wave state~\cite{rf:hasegawa,rf:won,
rf:zhitomirski,rf:izawa,rf:tanatar} requires a strong $k_{{\rm z}}$-dependence 
of the effective interaction. 
 This case is not expected in the quasi-2D systems like 
${\rm Sr}_{2}{\rm RuO}_{4}$.~\footnote{We estimate the three-dimensionality 
enough to give rise to the 3D $f$-wave state. The unrealistic value  
$t_{\perp}/t_{\parallel} \sim 0.1$ is required in the anisotropic cubic 
lattice.~\cite{rf:yanase3D}}

 The following part is constructed as follows. 
 In \S2, we explain the perturbation theory without spin-orbit interaction. 
 The formulation is given in \S2.1. 
 The basic results including the ODS are explained in \S2.2 and \S2.3. 
 In \S2.4, the mechanism of the superconductivity and the justification of 
the perturbation theory are discussed. 
 The effect of the spin-orbit interaction is taken into account in \S3. 
 The formulation is explained in \S3.1. 
 The results of the perturbation theory are shown in \S3.2. 
 The chiral state is stabilized when the $\gamma$-band is mainly 
superconducting. 
 The comparison with the other paring states is carried out in \S4.   
 \S5 is dedicated to the discussions on the validity of the approximations. 
 The conclusion and some discussions are given in \S6.

\section{Pairing Theory for the Three Band Model}

 In this section, we discuss the spin triplet superconductivity 
in the multi-band Hubbard model without spin-orbit interaction. 
 The perturbation theory is used to calculate the pairing 
interaction.~\cite{rf:nomura,rf:nomuramulti,rf:kohn,rf:hotta} 
 The results are qualitatively consistent with Refs. 21 and 22. 
 We explain the characteristic properties in the weak coupling region 
in more details. The robustness of the ODS is shown. 
 The validity of the perturbation theory and 
the comparison with the other calculations are discussed.

\subsection{Formulation}

 The three band model adopted for 
${\rm Sr_{2}RuO_{4}}$ is described as, 
\begin{eqnarray}
H &=& H_{0}+H_{{\rm LS}}+H_{{\rm I}}, 
\\
H_{0} &=& \sum_{a=1}^{3} \sum_{\k,s} \e_{a}(\k) 
c_{\mbox{{\scriptsize \boldmath$k$}},a,s}^{\dag} 
c_{\mbox{{\scriptsize \boldmath$k$}},a,s},  
\\
H_{{\rm LS}} &=& 2 \lambda \sum_{i} \LL_{i} \SS_{i}, 
\\
H_{{\rm I}} &=& 
U \sum_{i} \sum_{a} n_{i,a,\uparrow} n_{i,a,\downarrow} 
+ U' \sum_{i} \sum_{a>b} n_{i,a} n_{i,b} 
+ \Jh \sum_{i} \sum_{a>b} (2 \SS_{i,a} \SS_{i,b} + \frac{1}{2} n_{i,a} n_{i,b})
\nonumber \\
&&+ J \sum_{a \neq b} 
\sum_{\mbox{\boldmath$k$},\mbox{\boldmath$k'$},\mbox{\boldmath$q$}} 
c_{\mbox{{\scriptsize \boldmath$q$}}-\mbox{{\scriptsize \boldmath$k'$}},a,\downarrow}^{\dag} 
c_{\mbox{{\scriptsize \boldmath$k'$}},a,\uparrow}^{\dag} 
c_{\mbox{{\scriptsize \boldmath$k$}},b,\uparrow} 
c_{\mbox{{\scriptsize \boldmath$q$}}-\mbox{{\scriptsize \boldmath$k$}},b,\downarrow}, 
\\  
&=&
U \sum_{a} 
\sum_{\mbox{\boldmath$k$},\mbox{\boldmath$k'$},\mbox{\boldmath$q$}} 
c_{\mbox{{\scriptsize \boldmath$q$}}-\mbox{{\scriptsize \boldmath$k'$}},a,\downarrow}^{\dag} 
c_{\mbox{{\scriptsize \boldmath$k'$}},a,\uparrow}^{\dag} 
c_{\mbox{{\scriptsize \boldmath$k$}},a,\uparrow} 
c_{\mbox{{\scriptsize \boldmath$q$}}-\mbox{{\scriptsize \boldmath$k$}},a,\downarrow} 
\nonumber \\
&&+ U' \sum_{a>b} \sum_{s,s'} 
\sum_{\mbox{\boldmath$k$},\mbox{\boldmath$k'$},\mbox{\boldmath$q$}} 
c_{\mbox{{\scriptsize \boldmath$q$}}-\mbox{{\scriptsize \boldmath$k'$}},a,s}^{\dag} 
c_{\mbox{{\scriptsize \boldmath$k'$}},b,s'}^{\dag} 
c_{\mbox{{\scriptsize \boldmath$k$}},b,s'} 
c_{\mbox{{\scriptsize \boldmath$q$}}-\mbox{{\scriptsize \boldmath$k$}},a,s} 
\nonumber \\
&&+ \Jh \sum_{a>b} \sum_{s,s'} 
\sum_{\mbox{\boldmath$k$},\mbox{\boldmath$k'$},\mbox{\boldmath$q$}} 
c_{\mbox{{\scriptsize \boldmath$q$}}-\mbox{{\scriptsize \boldmath$k'$}},a,s}^{\dag} 
c_{\mbox{{\scriptsize \boldmath$k'$}},b,s'}^{\dag} 
c_{\mbox{{\scriptsize \boldmath$k$}},b,s} 
c_{\mbox{{\scriptsize \boldmath$q$}}-\mbox{{\scriptsize \boldmath$k$}},a,s'} 
\nonumber \\
&&+ J \sum_{a \neq b} 
\sum_{\mbox{\boldmath$k$},\mbox{\boldmath$k'$},\mbox{\boldmath$q$}} 
c_{\mbox{{\scriptsize \boldmath$q$}}-\mbox{{\scriptsize \boldmath$k'$}},a,\downarrow}^{\dag} 
c_{\mbox{{\scriptsize \boldmath$k'$}},a,\uparrow}^{\dag} 
c_{\mbox{{\scriptsize \boldmath$k$}},b,\uparrow} 
c_{\mbox{{\scriptsize \boldmath$q$}}-\mbox{{\scriptsize \boldmath$k$}},b,\downarrow}. 
\end{eqnarray}
 The three bands are constructed from the three $t_{{\rm 2g}}$-orbitals 
in the Ru ions. The index $a$ represents the 
$4d_{{\rm yz}}$ ($a=1$), $4d_{{\rm xz}}$ ($a=2$) and
$4d_{{\rm xy}}$ ($a=3$) orbitals, respectively. 
 The kinetic energy term $H_{0}$ is expressed by 
the two-dimensional tight-binding model, 
$\e_{1}(\k)=-2 t'_{{\rm xy}} \cos k_{{\rm x}} - 
2 t_{{\rm xy}} \cos k_{{\rm y}} - \mu_{{\rm xy}}$,  
$\e_{2}(\k)=-2 t_{{\rm xy}} \cos k_{{\rm x}} - 
2 t'_{{\rm xy}} \cos k_{{\rm y}} - \mu_{{\rm xy}}$ and 
 $\e_{3}(\k)=
-2 t_{{\rm z}} (\cos k_{{\rm x}} + \cos k_{{\rm y}}) - 4 
t'_{{\rm z}} \cos k_{{\rm x}} \cos k_{{\rm y}}  - \mu_{{\rm z}}$ 
with $t_{{\rm xy}} \gg t'_{{\rm xy}}$ and $ t_{{\rm z}} > t'_{{\rm z}}$. 
 Fig. 1 shows the typical Fermi surfaces 
which are consistent with the quantum oscillation 
measurement.~\cite{rf:mackenzie}  
 Hereafter, we describe the three bands without the spin-orbit interaction 
as $x$-, $y$- and $z$-band, respectively. 
 The $x$- and $y$-bands have a one-dimensional nature and 
construct the $\alpha$- and $\beta$-bands through a weak hybridization. 
 The weak hybridization is neglected for simplicity. 
 The $z$-band has a two-dimensional nature and corresponds 
to the $\gamma$-band.  
 Hereafter, we use the unit $\hbar=c=k_{{\rm B}}=1$ and choose the unit of 
energy as $t_{{\rm z}}=1$.

\begin{figure}[htbp]
  \begin{center}
    \includegraphics[height=6cm]{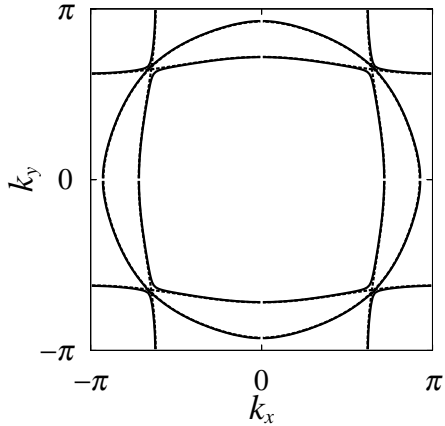}  
    \caption{Fermi surfaces with (solid lines) and without (dashed lines) 
             the spin-orbit coupling. Parameters are chosen 
             as $t'_{{\rm z}}=0.4$, $t_{{\rm xy}}=1.5$, $t'_{{\rm xy}}=0.2$ 
             and $n_{x}=n_{y}=n_{z}=1.33$. The spin-orbit coupling 
             is chosen as $\lambda=0.1$. 
             }
  \end{center}
\end{figure}

 The spin-orbit interaction in the ${\rm Ru}$ ions are expressed by 
$H_{{\rm LS}}$, where $2 \lambda$ is the coupling constant. 
 Typical value is considered to be $\lambda \sim 0.1$ which is much smaller 
than the band width. 
 In this section, we consider the case $\lambda=0$.  
 
 The interaction term $H_{{\rm I}}$ represents the on-site Coulomb 
interactions including the intra-band Coulomb term $U$, 
inter-band Coulomb term $U'$, Hund coupling term $\Jh$ 
and pair hopping term $J$. 
 The parameters satisfy the relation $U > 0$, $U' > 0$, $\Jh <0$, and  
$U > U' > |\Jh| \sim |J|$ in the ordinary situation. 
 The parameter $J$ has an ambiguity in its definition which is determined 
by the relative phase between the orbitals. 
 The standard notation gives a real positive value of $J$.

 Generally speaking, unconventional superconductivity originates from the 
momentum dependence of the effective interaction 
$\frac{1}{2} \sum \Gamma(k,k',a,a',b,b',s,s') 
c_{k,a,s}^{\dag} 
c_{-k,a',s'}^{\dag} 
c_{-k',b',s'}
c_{k',b,s}
$. 
 Here, we have only to consider the diagonal part 
$V_{a,b}(k,k',s,s') = \Gamma(k,k',a,a,b,b,s,s')$ 
since the logarithmic singularity appears from it.~\footnote{We safely 
neglect the Cooper pairing between the different band.} 
 Because the first order terms in the perturbation series 
do not contribute to the anisotropic superconductivity, 
the lowest order term is in the second order. 
 In the following, we calculate all the second order terms with respect to 
$H_{{\rm I}}$ and the third order terms with the coefficient $U^{3}$. 
 This approximation is justified in the 
perturbative region $U', |\Jh|, |J| < U \leq W$, where $W$ is 
the band width $W= 8$ in our unit. 
 We will explain the reason for including the third order terms in \S2.4.

 In case of $\lambda=0$, the SU(2) symmetry in the spin space is conserved. 
Then, the effective interaction in the triplet channel is given by 
$V^{{\rm t}}_{a,b}(k,k') = V_{a,b}(k,k',s,s')|_{s=s'}$ regardless of the 
$d$-vector.  
 Our procedure gives 
\begin{eqnarray}
 V^{{\rm t}}_{a,b}(k,k') &=& V^{(2)}_{a,b}(k,k')
                               + V^{(3)}_{a}(k,k') \delta_{a,b}, 
\\
  V^{(2)}_{a,b}(k,k') &=& 2 J \Jh \chi_{ab}(k-k') \ \ \ \ \ 
                           ({\rm for} \ \ \  a \neq b), 
\\
 V^{(2)}_{a,a}(k,k') &=& -U^{2} \chi_{a}(k-k') 
 -(2 U^{'2}+2 U' \Jh +\Jh^{2}) \sum_{\bar{a}} \chi_{\bar{a}}(k-k'), 
\\ 
  V^{(3)}_{a}(k,k') &=& 2 U^{3} {\rm Re} [\sum_{q} G_{a}(k + q) 
 G_{a}(k' + q) (\chi_{a}(q)-\phi_{a}(q))], 
\end{eqnarray} 
 where $G_{a}(k)=({\rm i}\omega_{n} - \e_{a}(\k))^{-1}$ is 
the Matsubara Green function. The indices $k$, $k'$ and $q$ are the four 
momentum $k = (\k, {\rm i}\omega_{n})$ {\it etc.}, and the summation is 
defined as $\sum_{k} = \frac{T}{N} \sum_{\k,n}$. 
 The index $\bar{a}$ in eq.(8) means the orbital with $\bar{a} \neq a$. 
 The functions $\chi_{a}(q)$, $\chi_{ab}(q)$ and $\phi_{a}(q)$ are 
defined as, 
\begin{eqnarray}
  \chi_{a}(q) &=& -\sum_{k} G_{a}(k+q) G_{a}(k), 
\\
  \chi_{ab}(q) &=& -{\rm Re} [\sum_{k} G_{a}(k+q) G_{b}(k)], 
\\
  \phi_{a}(q) &=& \sum_{k} G_{a}(q-k) G_{a}(k).
\end{eqnarray}

 The superconducting phase transition is determined by the linearized 
Dyson-Gorkov equation,  
\begin{eqnarray}
    \Delta_{a} (k) = 
 - \sum_{b,k'} V^{{\rm t}}_{a,b} (k,k') |G_{b}(k')|^{2} \Delta_{b}(k'), 
\end{eqnarray}
 where $\Delta_{a} (k)$ is the anomalous self-energy on the $a$-band. 
 The anomalous self-energy is an order parameter of the superconductivity. 
 The momentum and orbital dependence of the anomalous self-energy represents 
the wave function of the Cooper pairs. 
 The transition temperature is actually obtained by 
solving the \eli equation, 
\begin{eqnarray}
 \lambda_{{\rm e}}   \Delta_{a} (k) = 
 - \sum_{b,k'} V^{{\rm t}}_{a,b} (k,k') |G_{b}(k')|^{2} \Delta_{b}(k').  
\end{eqnarray} 
 The maximum eigenvalue $\lambda_{{\rm e}}$ becomes unity 
($\lambda_{{\rm e}}=1$) at the critical point. 
 It is notable that the \eli equation is reduced to the three independent 
equations if the off-diagonal term $V^{(2)}_{a,b}$ ($\propto J\Jh$) vanishes. 
 The transition temperature is uniquely determined 
owing to the pair hopping term $J$. 

 Let us remark here the relation to the excitation gap below \Tc. 
 The \eli equation determines the momentum and orbital dependence of the 
anomalous self-energy at $T=T_{\rm c}$. 
 Below \Tc, the Bogoliubov quasi-particle's energy is obtained 
as $E_{a}(\k)=z_{a}(\k)\sqrt{\varepsilon_{a}(\k)^{2}+
\Delta_{a}^{\rm ex}(\k)^{2}}$ where the renormalization factor $z_{a}(\k)$ 
is given by $z_{a}(\k)^{-1}=1-
\partial{\rm Re}\Sigma_{a}^{\rm R}(\k,\omega)/\partial\omega|_{\omega=0}$ and 
$\Delta_{a}^{\rm ex}(\k)=|\Delta^{\rm R}(\k,E_{a}(\k))|$, with 
$\Delta^{\rm R}_{a}(k)$ being the analytic continuation of $\Delta_{a}(k)$. 
 If we consider the weak coupling case, {\it i.e.} 
$T_{\rm c}, \Delta_{a}^{\rm ex}(\k) \ll W$, 
an approximation $\Delta^{\rm R}(\k,E_{a}(\k)) \simeq 
\Delta^{\rm R}(\k,0) \simeq \Delta_{a}(\k,{\rm i}\pi T)$ is very precise 
around the Fermi surface. 
 Therefore, the momentum and orbital dependence of the excitation gap is 
obtained from $|\Delta_{a}(\k,{\rm i}\pi T)|$ except for the factor arising 
from $z_{a}(\k)$. Since we found that the $a$ and $\k$-dependence of 
$z_{a}(\k)$ is not outstanding in Sr$_2$RuO$_4$, the qualitative nature of 
the excitation spectrum below $T_{\rm c}$ is obtained from the \eli equation. 
 Strictly speaking, the momentum and orbital dependence of $\Delta_{a}(k)$ 
is deformed below \Tc. However, this deformation is usually 
small when $T_{\rm c} \ll W$~\cite{rf:leggett}.

\subsection{Order parameter and transition temperature}

 In this section, the parameters and particle numbers are fixed to 
$(t'_{{\rm z}},t_{{\rm xy}},t'_{{\rm xy}})=(0.4,1.5,0.2)$, 
$(U,U',\Jh,J)=(5,1.5,-1,1)$ and $n_{x}=n_{y}=n_{z}=1.33$.
 The $z$-band has about 57 \% of the total DOS
under the above parameters. 
 Using $V^{{\rm t}}_{a,b} (k,k')$, we obtain $T_{{\rm c}}$ and 
the corresponding $\Delta_{a}(k)$. 
 In the summations, we divide the first Brillouin zone into 
$128 \times 128$ lattice points and take 2048 Matsubara frequencies. 
 We have confirmed that the obtained results do not depend on 
the numerical details, qualitatively.

\begin{figure}[htbp]
  \begin{center}
    \includegraphics[height=7.5cm]{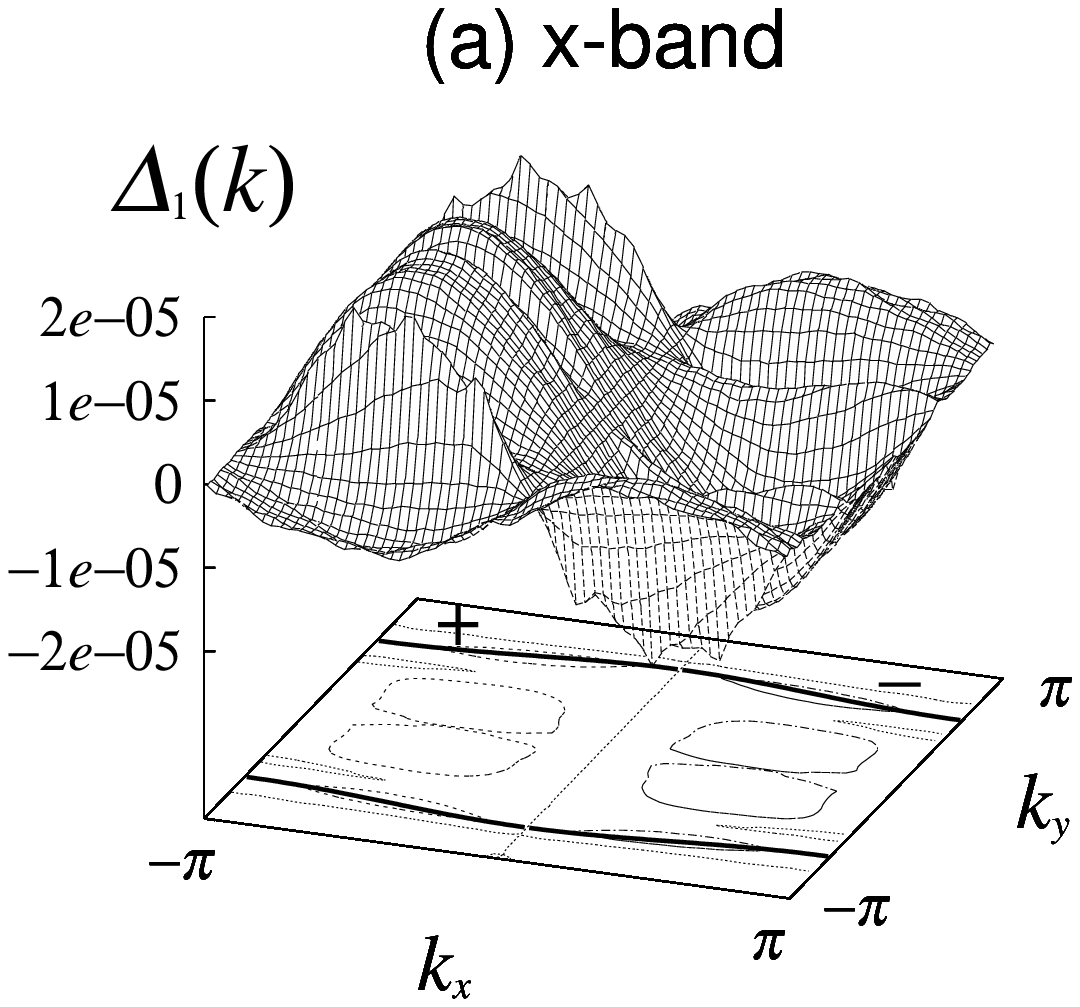}
    \includegraphics[height=7.5cm]{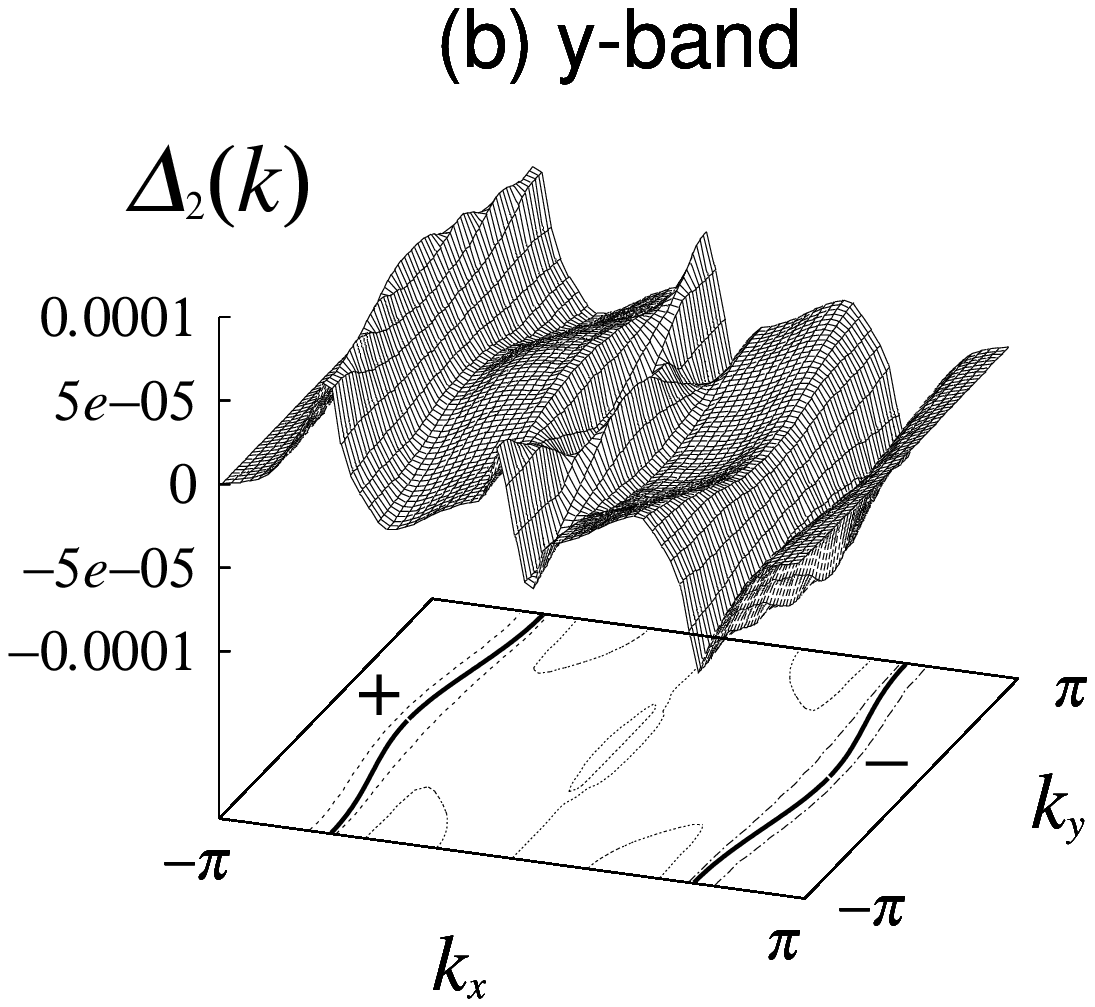}
    \includegraphics[height=7.5cm]{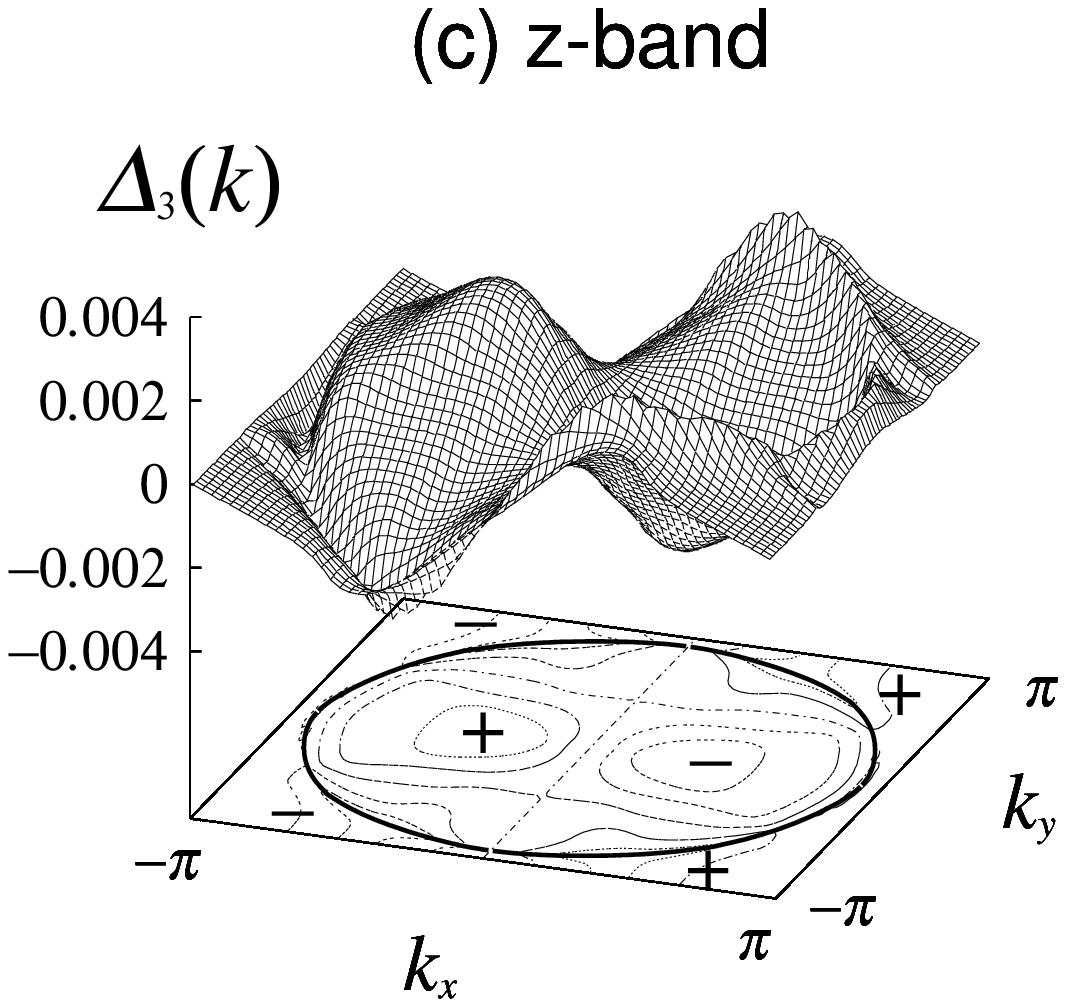}
    \caption{Momentum dependence of the order parameter 
             $\Delta_{a}(\k)=\Delta_{a}(\k,{\rm i}\pi T)$. 
             (a) $a=1$ ($x$-band), (b) $a=2$ ($y$-band) and (c) 
             $a=3$ ($z$-band). This figure shows the wave function  
             $\phi_{\rm x}(a,k)$. 
             The signs $+$ and $-$ show the sign of the 
             order parameter. The solid line in the basal plane shows 
             the Fermi surface in each band. 
             }
  \end{center}
\end{figure}

 First of all, let us study the symmetry of the order parameter. 
 The momentum and orbital dependence of 
$\Delta_{a}(\k)=\Delta_{a}(\k,{\rm i}\pi T)$ is shown 
in Fig. 2(a-c), where the temperature is chosen as $T=T_{{\rm c}}=0.0075$. 
 We find that the largest component is on the $z$-band ($\gamma$-band). 
 The pairing symmetry is the $p$-wave, which is consistent with 
Refs. 21 and 22.~\footnote{ 
In this paper, the pairing symmetry is defined by the sign reversal of 
the order parameter on the Fermi surface.} 
 The two nodes along $k_{{\rm x}}=0$ exist in the $x$ and $z$-bands. 
 There is no node in the $y$-band since the line $k_{{\rm x}}=0$ does not 
cross the Fermi surface. 
 The mechanism of stabilizing the triplet superconductivity will be discussed 
in \S2.4.

 From the symmetry, there is another solution $\phi_{{\rm y}}(a,k)$ 
which is rotated $90$ degrees from Fig. 2. 
 Generally speaking, arbitrary linear combinations 
$\Delta_{a}(k) = p \phi_{{\rm x}}(a,k) + q \phi_{{\rm y}}(a,k)$ 
are degenerate at $T=T_{\rm c}$. 
 We have shown the functions  
$\Delta_{a}(k) = \phi_{{\rm x}}(a,k)$ in Fig. 2. 
 The isotropic state like 
$\Delta_{a}(k) = \phi_{{\rm x}}(a,k) \pm {\rm i}\phi_{{\rm y}}(a,k)$ 
is expected below $T_{{\rm c}}$ in order to gain the condensation 
energy.~\cite{rf:leggett}

\begin{figure}[htbp]
  \begin{center}
    \includegraphics[height=6cm]{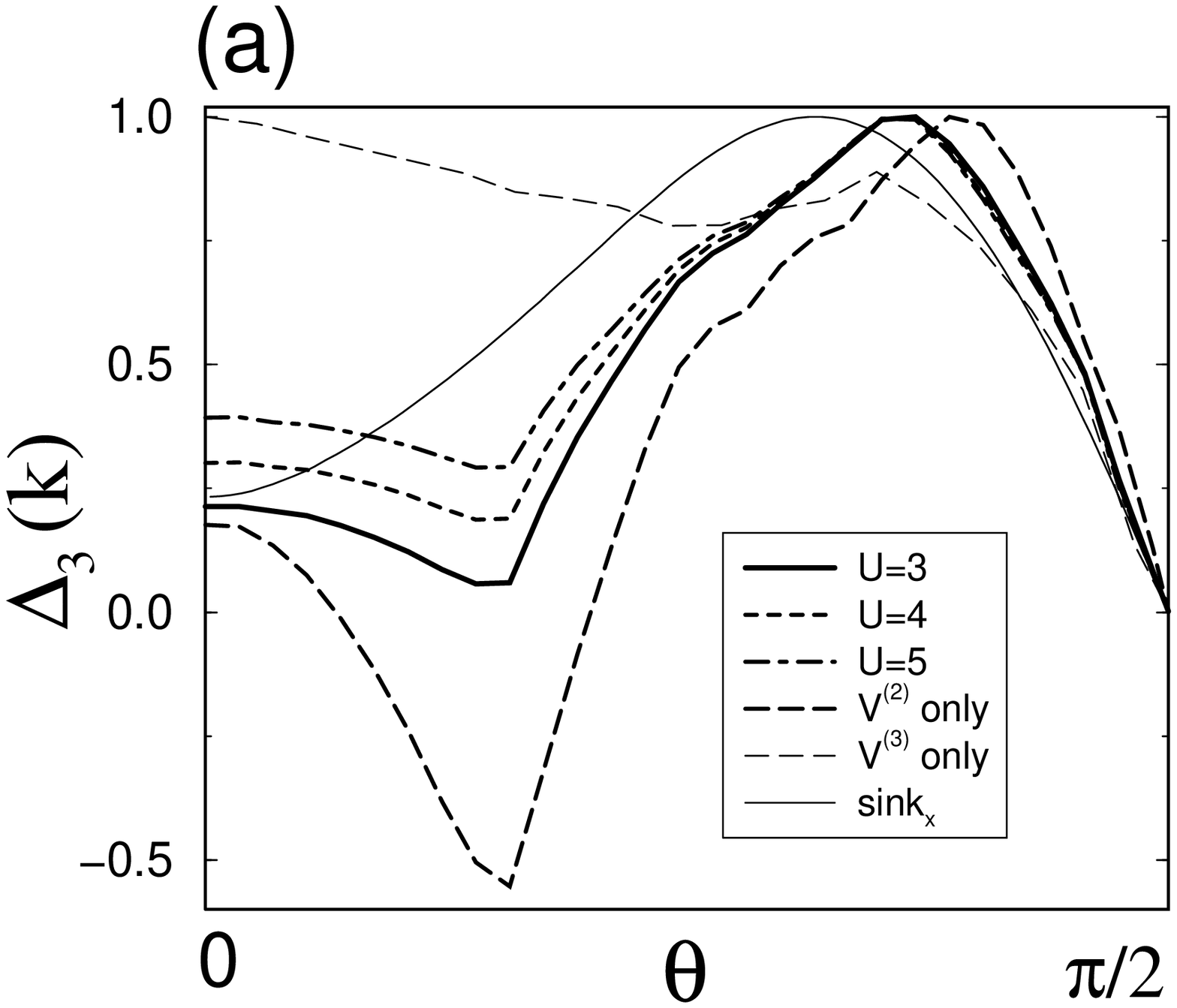}
    \hspace{10mm}
    \includegraphics[height=3cm]{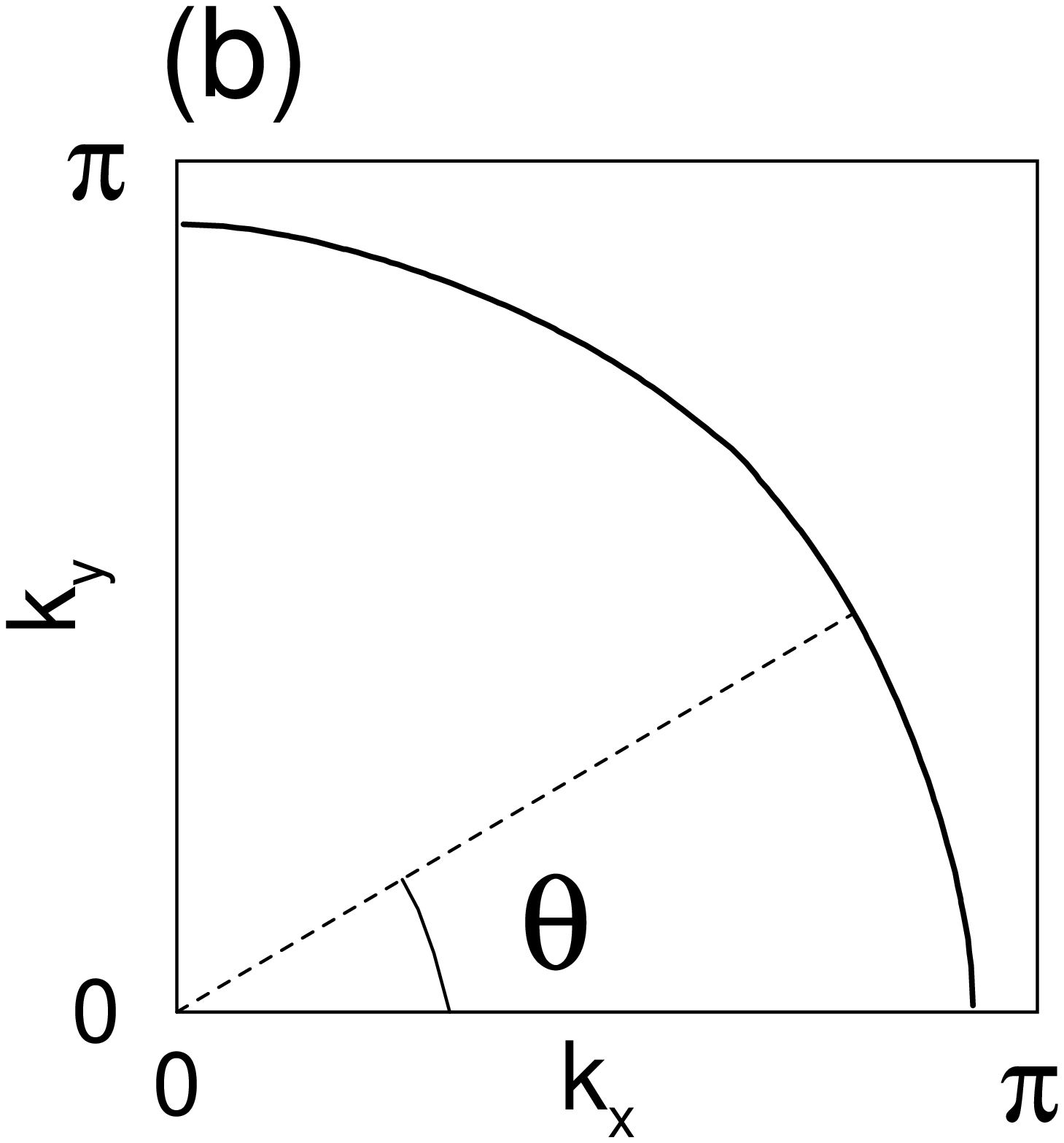}
    \caption{(a) Momentum dependence of the order parameter $\Delta_{3}(\k)$ 
             on the Fermi surface. The horizontal axis is the angle 
             $\theta={\rm Arctan}(k_{{\rm y}}/k_{{\rm x}})$ which is shown 
             in (b).  
             The results of the second order perturbation 
             (long-dashed), and the third order perturbation at 
             $U=3$ (thick solid), $U=4$ (dashed) and $U=5$ (dash-dotted) 
             are shown, respectively. 
             Here, the inter-band interactions are chosen as $U'=0.3 U$ and 
             $|\Jh|=J=0.2 U$. 
             The thin long-dashed line shows the result 
             when only the third order terms are taken into account. 
             The thin solid line shows the momentum dependence 
             $\Delta_{3}(\k)=\sin k_{{\rm x}}$. 
             The wave function is normalized as the maximum value is unity
             ($|\Delta_{3}(\k)|_{{\rm max}}=1$). 
            }    
  \end{center}
\end{figure}

 Second, let us see the angular dependence of the order parameter 
more closely. 
 The amplitude of $\Delta_{3}(\k)$ along the Fermi surface is shown in 
Fig. 3. 
 Apparently $\Delta_{3}(\k)$ is small around $(\pi,0)$, 
which is qualitatively consistent with 
the proposal by Miyake and Narikiyo.~\cite{rf:miyake} 
 Although this behavior may be important for the gap-less behaviors observed 
experimentally, we should point out that the excitation gap in the chiral 
state 
is not so tiny as to explain the power-law behaviors.

 We would like to stress that the momentum and frequency dependence of 
$\Delta_{a}(k)$ is determined so as to be most favorable for the 
superconductivity. 
 In other words, the wave function of the Cooper pairs is spontaneously 
distorted so that the attractive interaction works most efficiently. 
 It is interesting that the effective interaction $V^{{\rm t}}_{a,b}(k,k')$ 
(eqs.(6-9)) is repulsive for the simplified wave function 
$\Delta_{a}(k) \propto \sin k_{{\rm x}}$, but attractive for the distorted 
order parameter (shown in Fig. 2). 
 Thus, the strong coupling theory based on the \eli equation is essential 
for stabilizing the superconductivity.

\begin{figure}[htbp]
  \begin{center}
    \includegraphics[height=6cm]{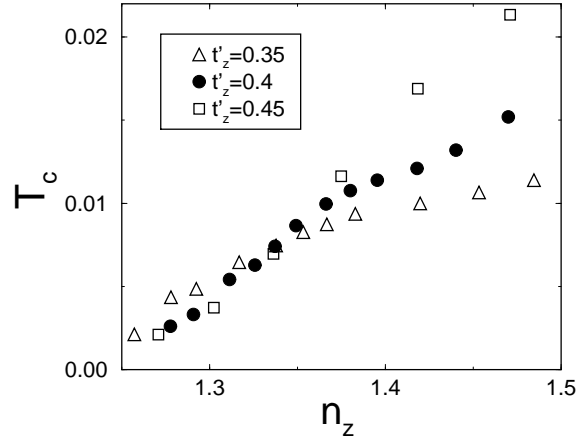}
    \caption{Transition temperature of the triplet 
             superconductivity for $t'_{{\rm z}}=0.35$ (triangle), 
             $t'_{{\rm z}}=0.4$ (circle)
             and $t'_{{\rm z}}=0.45$ (square).
             }
  \end{center}
\end{figure}

 Third, we discuss the parameter dependences of \Tc. 
 Figure 4 shows the $n_{z}$- and $t'_{{\rm z}}$-dependence. 
 Apparently, $T_{{\rm c}}$ increases with $n_{z}$. 
 This trend is consistent with the recent experiment.~\cite{rf:okuda} 
 We find that $T_{{\rm c}}$ is not sensitive to the Van-Hove 
singularity.~\cite{rf:comment21}  
 The triplet superconductivity with $T_{{\rm c}} \sim 0.007$ is obtained 
around $n_{z}=1.33$ irrespective of the value $t'_{{\rm z}}$. 
 If we assume $W = 2 {\rm eV}$, this critical temperature corresponds to 
$T_{{\rm c}} \sim 20 {\rm K}$. 
 We have chosen the parameters where $T_{{\rm c}}$ is higher than 
the experimental value $\sim 1.5 {\rm K}$ in order to keep the 
numerical accuracy. 
 Actually, $T_{{\rm c}} \sim 1.5 {\rm K}$ is obtained around $U=3.5$ 
(see \S2.4). 
 We have confirmed that the qualitative features discussed in this paper 
do not depend on the values $U$ and $T_{{\rm c}}$.

\subsection{Orbital dependent superconductivity}

 It is important to note that the amplitude of the order parameter 
remarkably depends on the orbitals. The maximum value of the order parameter 
$\Delta_{a}=|\Delta_{a}(k)|_{{\rm max}}$ is considerably small for $a=1,2$ 
(see Fig. 2). 
 Thus, the superconductivity is dominated by the $z$-band. 
 Such a strong orbital-dependence of the superconducting gap 
has been discussed phenomenologically~\cite{rf:agterberg} and called 
``orbital dependent superconductivity (ODS)''.  
 Hereafter, we denote the band in which the order parameter is largest as 
``main band'' and the other bands as ``minor bands''. 
 The main band is the $z$-band in the present case. This is mainly because the 
DOS is largest in the $z$-band.

\begin{figure}[htbp]
  \begin{center}
    \includegraphics[height=6cm]{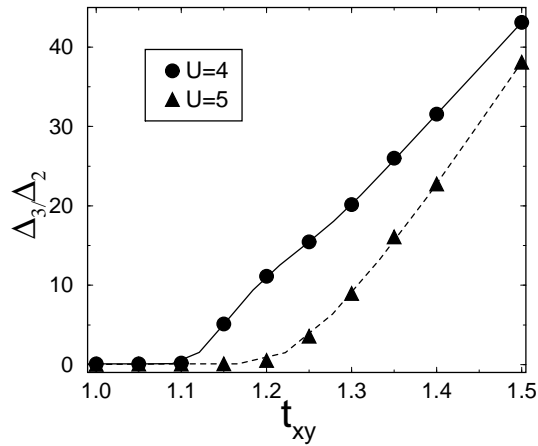}
    \caption{Ratio $\Delta_{3}/\Delta_{2}$ for   
             $U=4$ (circles) and $U=5$ (triangles). 
            }    
  \end{center}
\end{figure}

 We can see the robustness of the ODS behavior from Fig. 5 in which 
the $t_{{\rm xy}}$-dependence of the ratio $\Delta_{3}/\Delta_{2}$ is shown. 
 As $t_{{\rm xy}}$ decreases, the $y$-band becomes the main band since 
the DOS of the $y$-band increases. 
($\Delta_{1}$ is always small if the symmetry is fixed to 
$\Delta_{a}(k) = \phi_{{\rm x}}(a,k)$.) 
 Figure 5 shows that the decrease of $\Delta_{3}/\Delta_{2}$ is rapid, 
and the crossover region is very narrow. Thus, the order parameters 
in the different band {\it only accidentally} have the similar magnitude. 
 The parameter $t_{{\rm xy}}$ in the crossover region decreases 
with decreasing $U$.  
 We find that $z$-band is the main band in the weak coupling limit, 
when $t_{{\rm xy}} > 0.7$.

 If the hybridization between the orbitals is strong, we expect the similar 
amplitudes of order parameters among the bands. 
 In the present case, however, the coupling is only through the inter-band 
interaction $V_{a,b} \propto J \Jh$ which is much smaller than the intra-band 
one under the realistic condition. In other words, each band retains 
a character of the local orbital, and as a result, the ODS appears. 
 This fact will be the basis of the calculation in \S3 where we assume that 
the intra-band interaction is most important. 
 Even if the inter-band mixing between the $x$- and $y$-band is taken into 
account, the ODS behavior appears because the hybridization between 
the $z$-band and the other band vanishes in the 2D limit. This feature is 
robust owing to the inversion symmetry about the 
${\rm RuO}_{2}$-plane.~\cite{rf:agterberg}

\subsection{On the mechanism of the superconductivity}

 From the present results for $\lambda=0$, we discuss the mechanism of the 
superconductivity. 
 The following comment will justify the perturbation theory in this paper.

 Firstly, we wish to stress that the superconductivity is obtained 
within the second order perturbation, {\it i.e.}, in the lowest order 
perturbation. 
 Furthermore, the results of the $d$-vector discussed in \S3 do not 
change in the lowest order theory, qualitatively. 
 Therefore, the obtained results are not the artificial ones in the third 
order theory.

 However, $T_{{\rm c}}$ decreases exponentially in the small $U$ region  
and thus we need to choose moderate values of $U$ in order to obtain 
the experimental value $T_{{\rm c}} \sim 1.5{\rm K}$. 
 For example, $T_{{\rm c}} \sim 1.5{\rm K}$ is obtained around $U = 3.5$, 
$U'=0.3 U$ and $|\Jh|=J=0.2 U$,~\footnote{
Here, $T_{{\rm c}}$ is determined by the extrapolation, 
because our numerical calculation is justified above 
$T \geq 0.002$ ($\sim 6 {\rm K}$).} 
where the expansion parameter for the perturbation theory is $U/W \sim 0.38$.
 We have confirmed that the absolute value of the third order term 
$|V^{(3)}(k,k')|$ is still small in this region. For example, 
the average value of the third order term is about 
$|V^{(3)}| \sim 0.4 |V^{(2)}|$ in the above parameter set. 
 (Here, we have defined the average value as 
$|V^{(n)}|=\int_{{\rm F}} \int_{{\rm F}} 
|V_{3,3}^{(n)}(k,k')| {\rm d}k {\rm d}k'$, 
where $\int_{{\rm F}} {\rm d}k$ means the integration along the Fermi surface.)
 Therefore, we expect that the perturbation theory is qualitatively justified 
in this region.

 From the analysis below, we conclude that the role of the third order term is 
to stabilize the triplet superconductivity, {\it i.e.}, significantly enhance 
the eigenvalue $\lambda_{{\rm e}}$. 
 This is because the momentum dependence is well suitable 
for the triplet superconductivity rather than the second order term. 
 Therefore, the third order term plays an important role for the 
superconductivity in this region, which is the reason why we take into account 
the third order terms.

 To clarify the role of the third order terms, we compare the momentum 
dependence of $\Delta_{3}(\k)$ in Fig. 3 for various cases.  
 If we consider only the second order terms, the order parameter on the 
$z$-band has a higher symmetry; it has $10$ nodes on the Fermi surface. 
 However, this behavior will be fictitious because, when the third order terms 
are introduced, the pairing symmetry becomes $p$-wave. 
 This change is continuous, and the order parameter has similar form 
to the second order perturbation in the moderately weak coupling region. 
 On the other hand, the qualitatively different form is obtained when  
only the third order terms are considered. 
 These behaviors are natural because the system is in the perturbative region.

\begin{figure}[htbp]
  \begin{center}
    \includegraphics[height=6cm]{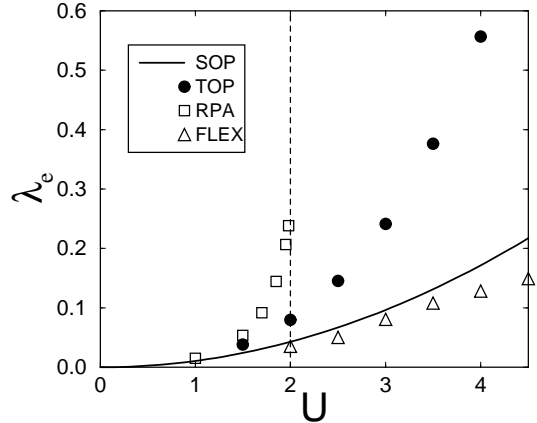}
    \caption{The comparison of $\lambda_{{\rm e}}$ in various calculations 
             for the triplet superconductivity. 
             The solid line, circles, squares and triangles show the results 
             of the second order perturbation (SOP), third order perturbation 
             (TOP), RPA and FLEX, respectively. 
             Here, the temperature is fixed to $T=0.005$ ($\sim 15 {\rm K}$) 
             where $\lambda_{{\rm e}} \sim 0.4$ corresponds to 
             the experimental value of $T_{{\rm c}} \sim 1.5 {\rm K}$ 
             by the extrapolation. 
             The parameters are chosen as $U'=\Jh=J=0$ for simplicity. 
             The dashed line shows the anti-ferromagnetic phase boundary 
             $U_{{\rm c}}=2.0$ in the mean field theory. 
             }    
  \end{center}
\end{figure}

 Let us discuss here the results obtained in the perturbative calculations, 
RPA and fluctuation exchange (FLEX) approximation. 
 In Fig. 6, we compare the eigenvalue $\lambda_{{\rm e}}$ in the 
various calculations for the triplet superconductivity. 
 In the RPA, the observed $T_{{\rm c}} \sim 1.5 {\rm K}$ 
is obtained only in the just vicinity of the magnetic order $U=2$. 
 This is mainly because the RPA term is not so suitable to the triplet pairing 
and the magnetic order is significantly overestimated in the RPA. 
 It is also important that the contribution from the magnetic fluctuation 
to the triplet pairing is $\frac{1}{3}$ of that to the 
singlet pairing.~\cite{rf:arita,rf:monthoux} 
 This fact can be remedied if the spin-orbit interaction is taken 
into account,~\cite{rf:ogata,rf:kuwabara,rf:satoh,rf:kuroki} 
but this issue is out of the scope of the present work.

 The magnetic order is actually suppressed by the fluctuations. 
 In Fig. 6, we also show the results of the FLEX 
approximation~\cite{rf:flex,rf:arita} which is a convenient calculation 
to include the fluctuations~\cite{rf:yanasereview}. 
 We can see that the eigenvalue $\lambda_{{\rm e}}$ does not easily develop 
in the FLEX approximation. 
 This is mainly because the de-pairing effect is very strong 
when the spin fluctuations have a local nature 
(namely, nearly $\q$-independent).~\footnote{
We should point out that the de-pairing effect is generally overestimated 
in the spin fluctuation theory.~\cite{rf:hotta,rf:yanasereview} 
This overestimation is serious when the spin fluctuations are spread 
in the momentum space. 
} 
 Since the spin fluctuations arising from the $z$-band have a local nature, 
the spin fluctuation mechanism is generally difficult for the $z$-band. 
 Actually, we find that the $x$- or $y$-band is the main band in the FLEX 
approximation for $U > 2.5$, where the pairing symmetry is the $f$-wave.

 These comparisons with the spin fluctuation theory indicate the 
importance of the non-RPA term which first appears in the third order 
correction. Actually, the third order terms significantly contribute to the 
enhancement of $\lambda_{{\rm e}}$ at $U=3$, although their absolute value 
is still small. It is not easy to guess what happens in the higher order 
perturbation series. 
 The actual result probably lies between the TOP and SOP results. 
 In any case the triplet superconductivity is possible out of the on-site 
repulsive interactions, and the weak coupling treatment is meaningful.

\begin{figure}[htbp]
  \begin{center}
    \includegraphics[height=6cm]{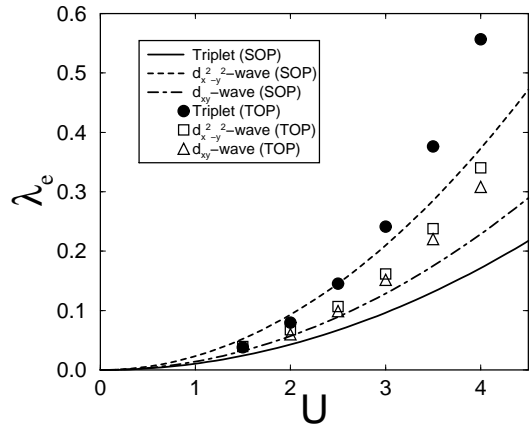}
    \caption{
             The comparison of $\lambda_{{\rm e}}$ for various symmetry. 
             The symbols are the results of the TOP for triplet (circle), 
             $d_{x^{2}-y^{2}}$-wave (square) and 
             $d_{xy}$-wave superconductivity (triangle). 
             The lines show the results of the SOP.
             The temperature is fixed to $T=0.005$. 
             The other parameters are chosen as $U'=0.3 U$ and $|\Jh|=J=0.2 U$.
             }    
  \end{center}
\end{figure}

 Before closing this section, let us compare with the possible singlet 
superconductivity. 
 Figure 7 similarly shows the eigenvalue $\lambda_{{\rm e}}$ at $T=0.005$.
 The results for the triplet, $d_{x^{2}-y^{2}}$-wave and $d_{xy}$-wave 
superconductivity are shown. 
 The largest value of $\lambda_{{\rm e}}$ indicates that the corresponding 
state is stabilized in the lower temperature. 
 If we consider only the second order perturbation, the $d_{x^{2}-y^{2}}$-wave 
symmetry is most stable. 
 However, the third order term significantly enhances the triplet 
superconductivity, and slightly suppresses the $d_{x^{2}-y^{2}}$-wave 
superconductivity. 
 We find that, although the RPA terms exist in the third order in the singlet 
pairing interaction $V^{{\rm s}}_{a,b}(k,k')$, they are considerably canceled 
by the non-RPA terms. This is an expected result when the spin fluctuation has 
a local nature.~\cite{rf:yanasereview} 
 As a result, the contribution from the third order terms to the 
$d_{x^{2}-y^{2}}$-wave pairing is small and negative. 
 On the contrary, the third order non-RPA terms significantly enhances 
$\lambda_{{\rm e}}$ for the triplet pairing. 
 Therefore, the triplet superconductivity is stabilized 
in the moderately weak coupling region ($U \geq 1.5$).

 The situation is very different for the case of High-$T_{{\rm c}}$ 
cuprates. The cancellation in the third order is not strong 
in the $d_{x^{2}-y^{2}}$-wave channel because of the nesting property of 
the Fermi surface or the strong tendency to the 
anti-ferromagnetism.~\cite{rf:yanasereview} 
 Then, it is generally expected that the contribution from the non-RPA terms 
is not severe. 
 Consequently, the spin fluctuation theory is qualitatively justified and 
the $T_{{\rm c}}$ for the $d_{x^{2}-y^{2}}$-wave superconductivity 
becomes very large.

\section{Determination of the $D$-vector}

 In this section, we determine the $d$-vector assuming that 
the most effective spin-orbit interactions are those in the Ru ions. 
 The SU(2) symmetry in the spin space is violated by this interaction, 
and thus the six-fold degeneracy of the triplet superconductivity is lifted. 
 We clarify which pairing state is stabilized by the spin-orbit interaction.

\subsection{Formulation}

 First, we explain the formulation to take into account the spin-orbit 
interaction $H_{{\rm LS}}$. 
 In this paper, $H_{{\rm LS}}$ is included in the unperturbed Hamiltonian 
$H'_{0}$ as  
\begin{eqnarray}
   H & = & H'_{0}+H_{{\rm I}},
\\
   H'_{0} & = &  H_{0} + H_{{\rm LS}}. 
\end{eqnarray}
 Hereafter, we choose $\lambda = 0.1$ unless we specify. 
 The results do not depend on the choice as long as $\lambda$ is much smaller 
than the band width. 
 By restricting the Hilbert space to the $t_{{\rm 2g}}$-orbitals, the 
unperturbed Hamiltonian is written in the following 
way,~\cite{rf:ngls,rf:ogata} 
\begin{eqnarray}
     H'_{0} = \sum_{\k,s} 
\left(
\begin{array}{ccc}
c_{\mbox{{\scriptsize \boldmath$k$}},1,s}^{\dag} &
c_{\mbox{{\scriptsize \boldmath$k$}},2,s}^{\dag} &
c_{\mbox{{\scriptsize \boldmath$k$}},3,-s}^{\dag}\\
\end{array}
\right)
\left(
\begin{array}{ccc}
\varepsilon_{1}(\k) & -s \lambda & -s \lambda\\
-s \lambda & \varepsilon_{2}(\k) &  \lambda\\
-s \lambda &  \lambda & \varepsilon_{3}(\k)\\
\end{array}
\right)
\left(
\begin{array}{c}
c_{\mbox{{\scriptsize \boldmath$k$}},1,s}\\
c_{\mbox{{\scriptsize \boldmath$k$}},2,s}\\
c_{\mbox{{\scriptsize \boldmath$k$}},3,-s}\\
\end{array}
\right). 
\end{eqnarray}
 The sign $s=1$ ($s=-1$) corresponds to the up (down) spin. 
 We denote the $3 \times 3$ matrix in eq.(17) as $\hat{H'_{0}}(\k,s)$. 
 Here, the wave functions of the $t_{{\rm 2g}}$-orbitals have been chosen as 
$4d_{{\rm yz}} = {\rm i} (4d_{1}+4d_{-1})/\sqrt{2}$, 
$4d_{{\rm xz}} = -{\rm i} (4d_{1}-4d_{-1})/\sqrt{2}$ and  
$4d_{{\rm xy}} = -{\rm i} (4d_{2}-4d_{-2})/\sqrt{2}$, respectively.  
 Note that a constant phase factor is multiplied to the $4d_{{\rm xz}}$ 
orbital in order to simplify the notation. 
 This definition makes the pair hopping term $J$ negative between the 
$4d_{{\rm xz}}$ and other orbitals.

 New quasi-particles are obtained by diagonalizing the matrix 
$\hat{H'_{0}}(\k,s)$ 
through the unitary transformation. 
 This procedure corresponds to the transformation of the basis as 
$
\left(
\begin{array}{ccc}
a_{\mbox{{\scriptsize \boldmath$k$}},1,s}^{\dag} &
a_{\mbox{{\scriptsize \boldmath$k$}},2,s}^{\dag} &
a_{\mbox{{\scriptsize \boldmath$k$}},3,s}^{\dag}\\
\end{array}
\right)
= \left(
\begin{array}{ccc}
c_{\mbox{{\scriptsize \boldmath$k$}},1,s}^{\dag} &
c_{\mbox{{\scriptsize \boldmath$k$}},2,s}^{\dag} &
c_{\mbox{{\scriptsize \boldmath$k$}},3,-s}^{\dag}\\
\end{array}
\right)
\hat{U}(\k,s), 
$
 where $\hat{U}(\k,s)=(u_{i,j}(\k,s))$ is a unitary matrix. 
 New quasi-particles created by the operators  
$a_{\mbox{{\scriptsize \boldmath$k$}},\alpha,s}^{\dag}$ are  
characterized by the pseudo-orbital $\alpha$ and pseudo-spin $s$. 
 The pseudo-spin corresponds not to the spin but to the Kramers doublet. 
 We can see that the rotational symmetry in the spin space is violated,  
and strictly speaking, the term ``odd-parity superconductivity'' should be 
used instead of the ``spin triplet superconductivity''. 
 Three Fermi surfaces of new quasi-particles have been shown in Fig. 1. 
 The effect of the spin-orbit coupling on the band structure is weak, 
except for the mixing around  
$\k=(\pm \frac{2}{3} \pi,\pm \frac{2}{3} \pi)$.

 The violation of the SU(2) symmetry typically appears in the spin 
correlation. We show the calculated results of the static 
spin susceptibility in the $\hat{z}$-direction 
$\chi_{{\rm z}}(\q)$ (Fig. 8(a)),  
and its difference from the planar component 
$\chi_{{\rm z}}(\q) - \chi_{\pm}(\q)$ (Fig. 8(b)). 
 We have used the Kubo formula and neglected the interaction term 
$H_{{\rm I}}$, for simplicity.   
 The static susceptibility has a peak around 
$\q=(\frac{2}{3}\pi,\frac{2}{3}\pi)$ which is consistent with the neutron 
scattering measurement~\cite{rf:sidis}. 
 The anisotropy results from the spin-orbit interaction. 
 The $\hat{z}$-component $\chi_{{\rm z}}(\q)$ is considerably 
larger around $\q=(\frac{2}{3}\pi,\frac{2}{3}\pi)$ and slightly 
smaller around $\q=(0,0)$. These properties 
are also consistent with the NMR measurements~\cite{rf:ishidaanisotropy} 
and the previous calculation~\cite{rf:ngmagne}.

\begin{figure}[htbp]
  \begin{center}
    \includegraphics[height=7cm]{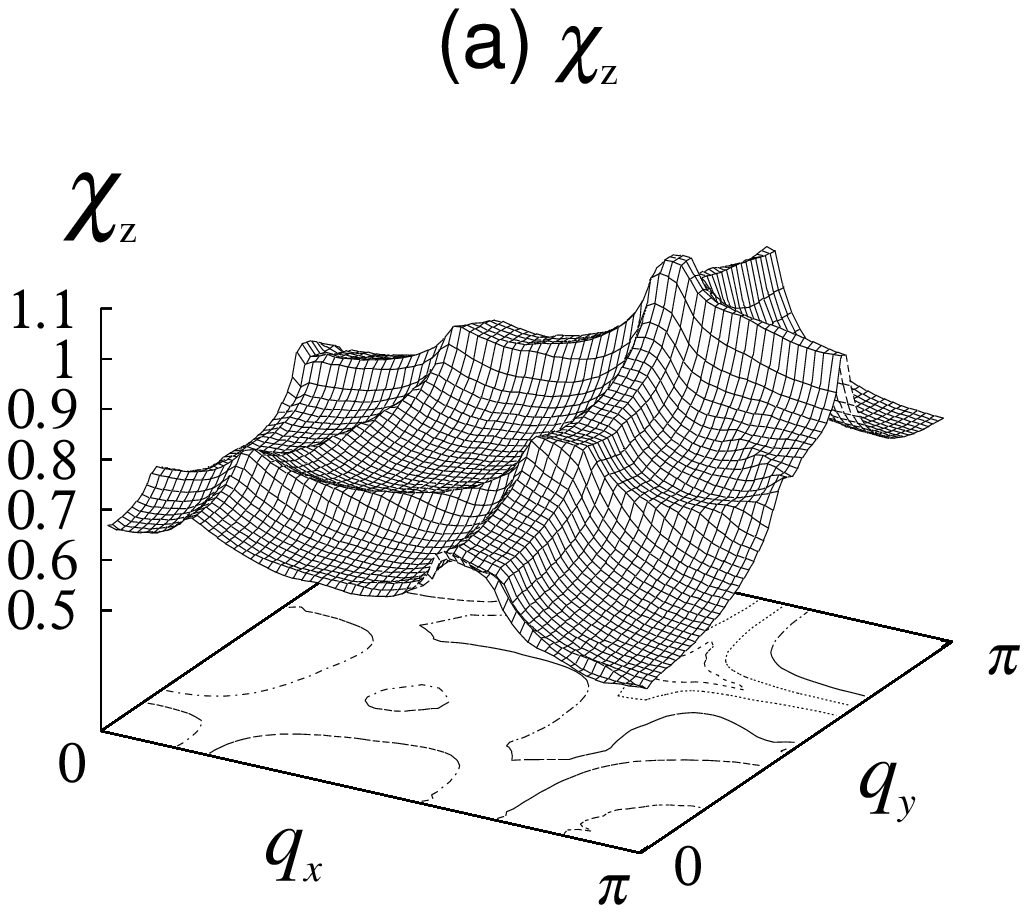} 
\hspace{10mm}
    \includegraphics[height=7cm]{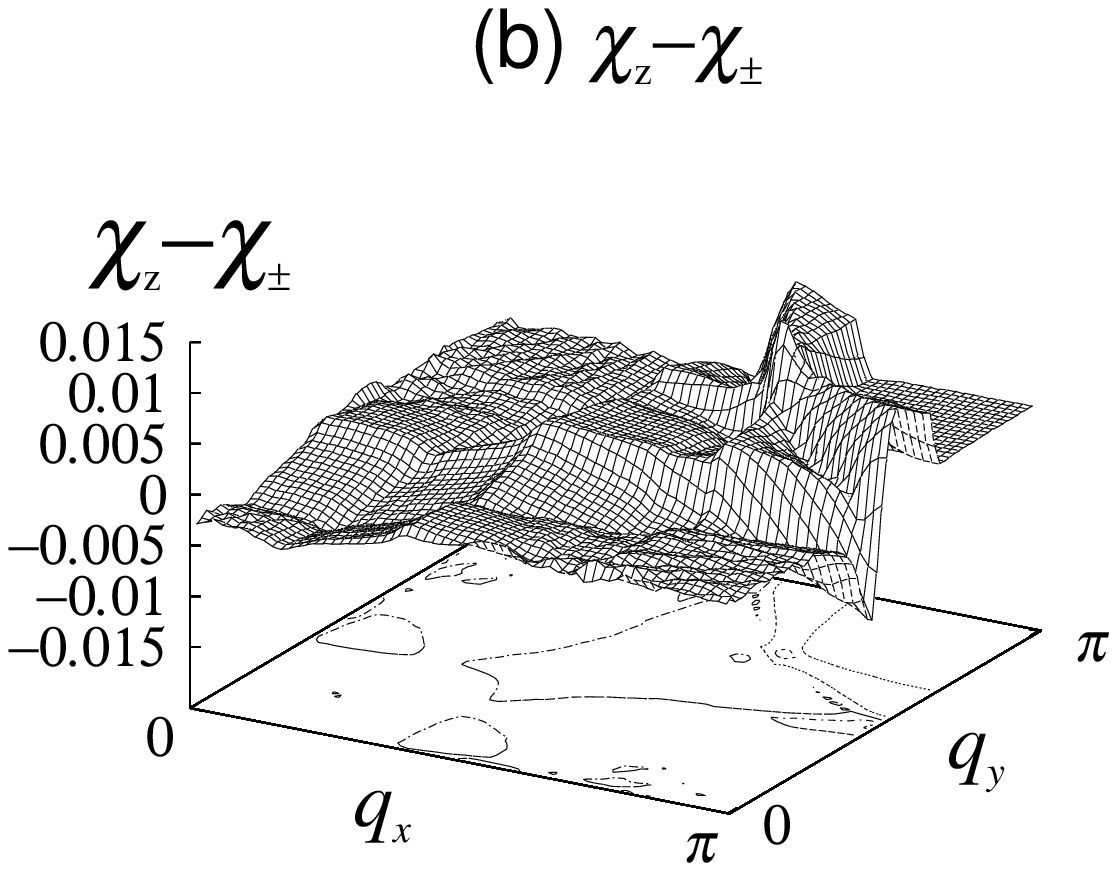} 
    \caption{(a) Static spin susceptibility of $z$-component 
             $\chi_{{\rm z}}(\q)$. (b) Anisotropy of the spin susceptibility 
             $\chi_{{\rm z}}(\q) - \chi_{\pm}(\q)$. 
             Here, $t'_{{\rm z}}=0.4$, $t_{{\rm xy}}=1.25$, 
             $t'_{{\rm xy}}=0.1$ and $\lambda=0.1$. 
             }
  \end{center}
\end{figure}

 We have to calculate the effective interactions between the new 
quasi-particles. 
 However, the interaction term $H_{{\rm I}}$ becomes very complicated 
by the unitary transformation. Therefore, we use the old basis 
$c_{\mbox{{\scriptsize \boldmath$k$}},a,s}$ tentatively, and at the end 
we apply the unitary transformation to obtain the effective interactions 
in the new basis. 
 The effective interaction $\Gamma$ between the old quasi-particles 
is calculated by the perturbative method within the same order as in \S2. 
 It should be noticed that the many terms are added to $\Gamma$ through 
the off-diagonal Green functions. Moreover, we have to calculate the 
off-diagonal part of $\Gamma$ because it contributes through the unitary 
transformation. 
 Finally, the effective interaction $\tilde{\Gamma}$ between the new 
quasi-particles is obtained as, 
\begin{eqnarray}
& & \tilde{\Gamma}(k,k',\alpha,\alpha',\beta',\beta,s_{1},s_{2},s_{3},s_{4}) =
\nonumber \\
& & \sum_{a,a',b,b'} 
u^{*}_{a,\alpha}(\k,s_{1}) 
u^{*}_{a',\alpha'}(-\k,s_{2}) 
\Gamma(k,k',a,a',b',b,s_{1},s_{2},s_{3},s_{4}) 
u_{b',\beta'}(-\k',s_{3}) u_{b,\beta}(\k',s_{4}), 
\end{eqnarray} 
and therefore, 
\begin{eqnarray}
& & \frac{1}{2} \sum_{a,a',b,b'} \sum_{s_{1},s_{2},s_{3},s_{4}} 
\Gamma(k,k',a,a',b',b,s_{1},s_{2},s_{3},s_{4}) 
c_{k,a,s_{1}}^{\dag} 
c_{-k,a',s_{2}}^{\dag} 
c_{-k',b',s_{3}}
c_{k',b,s_{4}} 
\nonumber \\
& & = 
\frac{1}{2} \sum_{\alpha,\alpha',\beta,\beta'} \sum_{s_{1},s_{2},s_{3},s_{4}} 
\tilde{\Gamma}(k,k',\alpha,\alpha',\beta',\beta,s_{1},s_{2},s_{3},s_{4}) 
a_{k,\alpha,s_{1}}^{\dag} 
a_{-k,\alpha',s_{2}}^{\dag} 
a_{-k',\beta',s_{3}}
a_{k',\beta,s_{4}}. 
\end{eqnarray}
 Here, we have redefined the up (down) spin in $z$-orbital as $s=-1$ ($s=1$). 
 The diagonal part about the pseudo-orbital 
$
\tilde{V}_{\alpha,\beta}(k,k',s_{1},s_{2},s_{3},s_{4}) 
=\tilde{\Gamma}(k,k',\alpha,\alpha,\beta,\beta,s_{1},s_{2},s_{3},s_{4}) 
$
is regarded as the pairing interaction. 
 As a result, the \eli equation is extended in the following way, 
\begin{eqnarray}
 \lambda_{{\rm e}}   \Delta_{\alpha,s_{1},s_{2}} (k) = 
 - \sum_{\beta,k',s_{3},s_{4}} 
\tilde{V}_{\rm \alpha,\beta} (k,k',s_{2},s_{1},s_{3},s_{4}) 
|\tilde{G}_{\beta}(k')|^{2} \Delta_{\beta,s_{3},s_{4}}(k').  
\end{eqnarray} 
 Here, $ \tilde{G}_{\alpha}(k) = ({\rm i}\omega_{n} - E_{\alpha}(\k))^{-1}$ 
is the Green function characterized by the pseudo-orbital where 
$E_{\alpha}(\k)$ is the energy of new quasi-particles.

 The practical calculation for the above equation is generally  
tedious since the pairing interaction 
$\tilde{V}_{\rm \alpha,\beta} (k,k',s_{1},s_{2},s_{3},s_{4})$ 
has so many terms.  
 However, we can simplify the calculation by using the additional 
approximation based on the ODS argument combined with the perturbation 
with respect to the spin-orbit coupling $\lambda$. 
 As is shown in \S2.3, the amplitude of the order parameter strongly depends 
on the orbital. In this case, the eigenvalue $\lambda_{{\rm e}}$ is almost 
determined by the interaction between the main orbital.  
 This feature is justified for the pseudo-orbital since the spin-orbit 
coupling is weak. 
 Therefore, it is sufficient to take into account only the diagonal part 
of the interaction 
$\tilde{V}_{\rm \alpha}(k,k',s_{1},s_{2},s_{3},s_{4})=\tilde{V}_{\rm \alpha,\alpha}(k,k',s_{1},s_{2},s_{3},s_{4})$, 
and investigate each case where the main pseudo-orbital is 
$\alpha=1,2$ or $3$. 
 The validity of this approximation will be discussed in \S5.

 The perturbation with respect to $\lambda$ is performed in the following way. 
 First, the Green function 
$\hat{G}(k,s) = ({\rm i}\omega_{n} \hat{1} - \hat{H'_{0}}(\k,s))^{-1}$
and the unitary matrix  $\hat{U}(\k,s)$ are expressed as, 
\begin{eqnarray}
\hat{G}(k,s)& = &
\left(
\begin{array}{ccc}
g_{2}(k) g_{3}(k) - \lambda^{2} 
& -s \lambda (g_{3}(k) + \lambda) 
&  -s \lambda (g_{2}(k) + \lambda) \\
-s \lambda (g_{3}(k) + \lambda) 
&g_{1}(k) g_{3}(k) - \lambda^{2} 
&  \lambda (g_{1}(k) + \lambda) \\
-s \lambda (g_{2}(k) + \lambda) 
& \lambda (g_{1}(k) + \lambda)
& g_{1}(k) g_{2}(k) - \lambda^{2}
\end{array}
\right)/A(k),
\\
A(k) &=& 
g_{1}(k) g_{2}(k) g_{3}(k) - \lambda^{2} (g_{1}(k) + g_{2}(k) + g_{3}(k))
-2 \lambda^{3}, 
\\
\hat{U}(\k,s)& = &
\left(
\begin{array}{ccc}
z_{{\rm x}}(\k)^{1/2} 
& -s \frac{\lambda}{\e_{y}(\mbox{{\scriptsize \boldmath$k$}})-\e_{x}(\mbox{{\scriptsize \boldmath$k$}})} 
& -s \frac{\lambda}{\e_{z}(\mbox{{\scriptsize \boldmath$k$}})-\e_{x}(\mbox{{\scriptsize \boldmath$k$}})}\\
-s \frac{\lambda}{\e_{x}(\mbox{{\scriptsize \boldmath$k$}})-\e_{y}(\mbox{{\scriptsize \boldmath$k$}})} 
& z_{{\rm y}}(\k)^{1/2} 
& \frac{\lambda}{\e_{z}(\mbox{{\scriptsize \boldmath$k$}})-\e_{y}(\mbox{{\scriptsize \boldmath$k$}})} \\
-s \frac{\lambda}{\e_{x}(\mbox{{\scriptsize \boldmath$k$}})-\e_{z}(\mbox{{\scriptsize \boldmath$k$}})} 
& \frac{\lambda}{\e_{y}(\mbox{{\scriptsize \boldmath$k$}})-\e_{z}(\mbox{{\scriptsize \boldmath$k$}})} 
& z_{{\rm z}}(\k)^{1/2} \\
\end{array}
\right),    
\end{eqnarray}
 where $g_{a}(k) = {\rm i}\omega_{n} - \e_{a}(\k)$.  
 The diagonal component $z_{a}(\k)^{1/2}$ is a normalization factor.  
 Equation (23) is a correct expression within the lowest order about 
$\lambda$, which is sufficient for the following calculation. 
 It is important that the off-diagonal component of the matrices 
(eqs.(21) and (23)) appears in the order of $\lambda$. 
 From this observation, we find that the lowest order correction in the 
pairing interaction $\tilde{V}_{\rm \alpha}(k,k',s_{1},s_{2},s_{3},s_{4})$ 
is in the second order. 
 Therefore, 
we perform the calculation up to $O(\lambda^{2})$ in the following. 
 This procedure is justified because $\lambda$ is sufficiently small 
in ${\rm Sr}_{2}{\rm RuO}_{4}$ ($\lambda/W \sim 0.02$). 
 The normalization factor $z_{a}(\k)=1 - O(\lambda^{2})$ also contributes 
within the second order.

 We find that the SU(2) symmetry in the $d$-vector space is violated by  
a partial set of the second order terms with respect to $\lambda$, which 
we call the ``symmetry breaking interaction''. 
 We find that the symmetry breaking interaction vanishes in case of $\Jh=0$. 
 That is, the SU(2) symmetry in the $d$-vector space is violated 
by the combination of the Hund coupling and the spin-orbit interaction. 
 Note that the SU(2) symmetry in the $d$-vector space can be conserved 
even if the SU(2) symmetry in the real spin is violated.

 As is shown in eq.(23), the expansion parameter is 
$\lambda/|\e_{a}(\k)-\e_{b}(\k)|$. 
 This is sufficiently small ($\sim \lambda/t_{{\rm z}}$) except for the 
crossing point of the Fermi surface around 
$\k \sim (\pm\frac{2}{3}\pi, \pm\frac{2}{3}\pi)$. 
 We have confirmed that the contribution from this region 
is negligible. 
 We have also confirmed that the energy of the new quasi-particles, 
$E_{\alpha}(\k)$, is safely approximated as that of the corresponding 
old quasi-particles $\e_{a}(\k)$ which is the main component of 
the pseudo-orbital $\alpha$. 
 This approximation does not affect on the results within the numerical error. 
 We define the $\alpha'$-, $\beta'$- and $\gamma$-band of the 
new quasi-particles which corresponds to the old $x$-, $y$- and $z$-band, 
respectively. 
 Strictly speaking, $\alpha'$ and $\beta'$ are not the band, but constructed  
from the part of the $\alpha$- and $\beta$-band. 
 Here, we have used the index representing the local orbital which is more 
convenient to discuss the superconductivity. 
 The ODS means the strong dependence of the order parameter 
on the $\alpha'$-, $\beta'$- and $\gamma$-band. 
 
 Finally, the eigenstates are classified by using the $d$-vector. 
 From the symmetry argument, we have the following eigenstates 
(1) $\hat{d}(k) = k_{{\rm x}} \hat{x} \pm k_{{\rm y}} \hat{y}$, 
(2) $\hat{d}(k) = k_{{\rm x}} \hat{y} \pm k_{{\rm y}} \hat{x}$ and  
(3) $\hat{d}(k) = (k_{{\rm x}} \pm {\rm i} k_{{\rm y}}) \hat{z}$.  
 Here, $k_{{\rm x}}$ and $k_{{\rm y}}$ represent the reflection symmetry 
of the wave function and does not mean the detailed momentum dependence. 
 There remains the two-fold degeneracy in each cases. 
 Although other linear combinations are possible, we have chosen  
the symmetric states (1)-(3), which  
are expected to be stabilized in order to gain the condensation energy. 
  The degeneracy in (1) and (2) is finally lifted by the weak mixing between 
the $x$- and $y$-orbitals, but the main results of this paper are not affected 
by this effect. 
 A brief comment on this point will be given in \S5.

 Because the symmetry $d_{n}(k)=\phi_{{\rm x}}(k)$ decouples with 
the different symmetry $d_{n'}(k)=\phi_{{\rm y}}(k)$ within our approximation, 
 the effective interaction is defined according to the direction 
of the $d$-vector ($n=x,y,z$), 
\begin{eqnarray}
\tilde{V}_{n,\alpha} (k,k') = \frac{1}{2}
({\rm i} \sigma_{n} \sigma_{{\rm y}})^{\dag}_{s_{1},s_{2}}
 \tilde{V}_{\alpha} (k,k',s_{1},s_{2},s_{3},s_{4})
({\rm i} \sigma_{n} \sigma_{{\rm y}})_{s_{3},s_{4}}. 
\end{eqnarray} 
 Here, the spin part of the order parameter is assigned by the $d$-vector 
as, 
\begin{eqnarray}
\left(
\begin{array}{cc}
\Delta_{\uparrow\uparrow}(k) & \Delta_{\uparrow\downarrow}(k) \\
\Delta_{\downarrow\uparrow}(k) & \Delta_{\downarrow\downarrow}(k) \\
\end{array}
\right)
=
\left(
\begin{array}{cc}
-d_{{\rm x}}(k)+{\rm i}d_{{\rm y}}(k) & d_{{\rm z}}(k) \\
d_{{\rm z}}(k) & d_{{\rm x}}(k)+{\rm i}d_{{\rm y}}(k) \\
\end{array}
\right)
=
{\rm i} \hat{d}(k) \hat{\sigma} \sigma_{y}.
\end{eqnarray}
  The resulting eigenvalue equation is obtained respectively, 
\begin{eqnarray}
 \lambda_{{\rm e}}   d_{n}(k) = 
 - \sum_{k'} \tilde{V}_{n,\alpha} (k,k') 
|G_{\alpha}(k')|^{2} d_{n}(k'). 
\end{eqnarray} 
 It should be again stressed that the effective interaction 
$ \tilde{V}_{n,\alpha}(k,k')$ depends on the 
main band $\alpha$ and the direction of the $d$-vector $n$.

 The numerical calculation is used to solve the three kinds of the \eli 
equations (eq.(26)), which correspond to the three kinds of eigenstates 
(1)-(3). 
 The three different transition temperatures are obtained when $\Jh \neq 0$. 
 We regard that the state with maximum $T_{{\rm c}}$ is stabilized. 
 This procedure is surely correct around $T \sim T_{{\rm c}}$. 
 Strictly speaking, the determination of the ground state $d$-vector requires 
a direct estimation of the condensation energy, which was not performed in 
this paper. 
 If the ground state $d$-vector is different from that for 
$T \sim T_{{\rm c}}$, a double transition is indicated. However, there is no  
apparent reason for the double transition in the following sense. 
 The condensation energy is maximumly gained in the above six eigenstates 
within the weak coupling theory for $\lambda=0$. 
 We can show that the quadratic term in the Ginzburg-Landau expansion 
stabilizes the chiral state at finite $\lambda$. 
 Unless the higher order terms have an especial behavior, 
the stabilized state is retained below \Tc. 
 Also we know that the double transition is not observed experimentally 
under the zero magnetic field.~\cite{rf:nishizaki,rf:deguchi}

\subsection{Results of the perturbation theory}

 First, we perform a fully microscopic calculation using the perturbation 
theory. 
 In this case, the superconductivity is the $p$-wave and the main band is 
$\gamma$ as is shown in \S2. 
 This case is most feasible, since the DOS is largest in the $\gamma$-band. 
 In this subsection, we show that the $d$-vector is also consistent 
with experiments.

 We calculate the diagonal part of the interaction between the $\gamma$-band, 
$\tilde{V}_{n,\gamma}(k,k')$ and solve the corresponding \eli equations 
(eq.(26)). 
 Figure 9 shows the $\lambda$-dependence of the transition temperature.  
 We can see that $T_{{\rm c}}$ is reduced by the spin-orbit coupling. 
 This is mainly because the contribution from the zeroth order 
term $V_{3,3}(k,k')$ 
monotonically decreases because of the normalization factor 
$z_{{\rm z}}(\k) z_{{\rm z}}(\kk) \sim 1 - O(\lambda^{2})$. 
 Although many terms arise from the spin-orbit interaction, they do not work 
as an attractive interaction. 
 We confirm that $T_{{\rm c}}$ at $\lambda=0$ coincides with the result 
in the three band model (Fig. 4) within the numerical error. 
 Thus, our approximation based on the ODS gives an appropriate estimation.

\begin{figure}[htbp]
  \begin{center}
    \includegraphics[height=6cm]{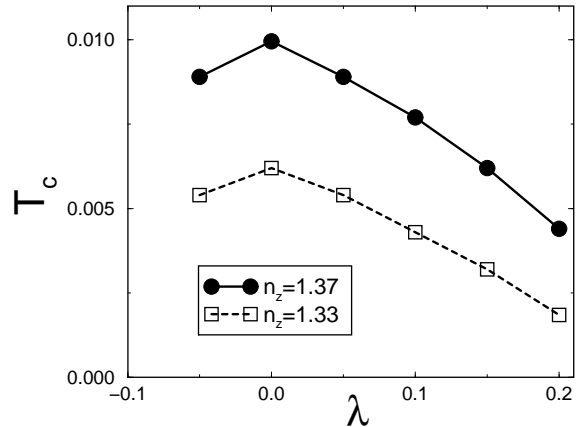} 
    \caption{$\lambda$-dependence of the transition temperature for 
             $n_{{\rm z}} = 1.33$ and $n_{{\rm z}} = 1.37$. 
             }
  \end{center}
\end{figure}

 The temperature dependence of the eigenvalue $\lambda_{{\rm e}}$ 
for three kinds of the eigenstates is shown in Fig. 10. 
 Hereafter, we choose the parameter set $t_{{\rm xy}}=1.25$ and 
$t'_{{\rm xy}}=0.1$ which increases the splitting of the degeneracy. 
 We can see that the chiral state 
(3) $\hat{d}(k) = (k_{{\rm x}} \pm {\rm i} k_{{\rm y}}) \hat{z}$ 
is stabilized in our calculation. 
 We will show that this result is robust for the reasonable parameter set.  
 Because the splitting of the eigenvalue $\lambda_{{\rm e}}$ is almost 
independent of the temperature, we determine the stabilized state as a state 
with maximum eigenvalue $\lambda_{{\rm e}}$ at the fixed temperature $T=0.01$, 
in the following.

 Note that the splitting of $T_{\rm c}$ is considerably small, 
{\it i.e.}, $\Delta T_{{\rm c}} \sim 0.01 T_{{\rm c}}$ in this case. 
 This is mainly because the symmetry breaking interaction is in 
the second order with respect to $\lambda$, and requires 
the Hund coupling term $\Jh$. 
 The symmetry breaking interaction typically has the order  
$\sim U \Jh \lambda^{2}/W^{3}$ which is much smaller than the zeroth order 
term $V_{3,3}(k,k') \sim U^{2}/W$.  
 Thus, the difference of the condensation energy is much smaller than the 
spin-orbit coupling energy $\lambda$. 
 Note that the spin-orbit coupling makes the spin correlation anisotropic 
even if $\Jh=0$ (Fig. 8). Therefore, it is generally expected that 
the violation of the SU(2) symmetry in the $d$-vector space is smaller than 
that in the real spin space. 
 This result may be connected with the multiple phase diagram of 
${\rm UPt}_{3}$.~\cite{rf:tou,rf:ohmi,rf:sauls} 
 A brief discussion on this point will be given in \S6.

\begin{figure}[htbp]
  \begin{center}
    \includegraphics[height=5.5cm]{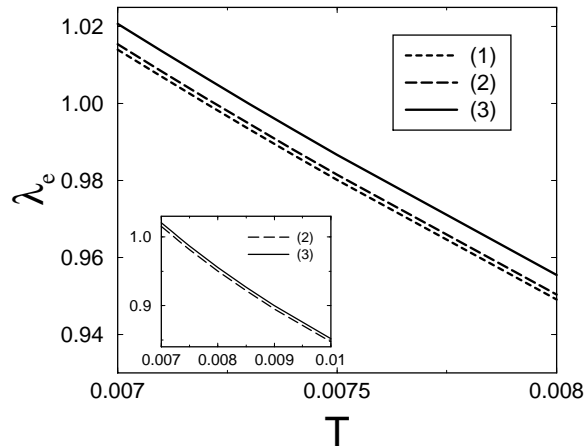} 
    \caption{Temperature dependence of $\lambda_{{\rm e}}$ for three kinds of 
             the eigenstates 
            (1) $\hat{d}(k) = k_{{\rm x}} \hat{x} \pm k_{{\rm y}} \hat{y}$ 
            (dashed-line), 
            (2) $\hat{d}(k) = k_{{\rm x}} \hat{y} \pm k_{{\rm y}} \hat{x}$ 
            (long dashed-line) and  
            (3) $\hat{d}(k) = (k_{{\rm x}} \pm {\rm i} k_{{\rm y}}) \hat{z}$ 
            (solid line). 
            We can see that the state (3) is stabilized by the spin-orbit 
            interaction. 
            The inset shows the same quantities in the wide temperature region.
             }
  \end{center}
\end{figure}

 The phase diagram is shown in Fig. 11, which is a main result of this paper. 
 The chiral state 
(3) $\hat{d}(k) = (k_{{\rm x}} \pm {\rm i} k_{{\rm y}}) \hat{z}$ 
is stabilized in the wide region around $n_{{\rm z}} \sim 1.33$. 
 We have confirmed that the result in Fig. 11 is robust for the parameters 
of the $\alpha'$- and $\beta'$-band, 
only if the particle number is around $n_{{\rm x}}= n_{{\rm y}} \sim 1.3$. 
 The other state (1) or (2) is stabilized only in the unphysical cases, 
for example, (i) hole-like Fermi surface of the $\gamma$-band (see Fig. 11), 
(ii) half-filled $\alpha'$- and $\beta'$-band,  
$n_{{\rm x}}= n_{{\rm y}} \sim 1$ or 
(iii) anti-ferromagnetic Hund coupling $\Jh > 0$. 
 Consequently, we conclude that the chiral state is robustly obtained in this 
calculation.

\begin{figure}[htbp]
  \begin{center}
    \includegraphics[height=6.2cm]{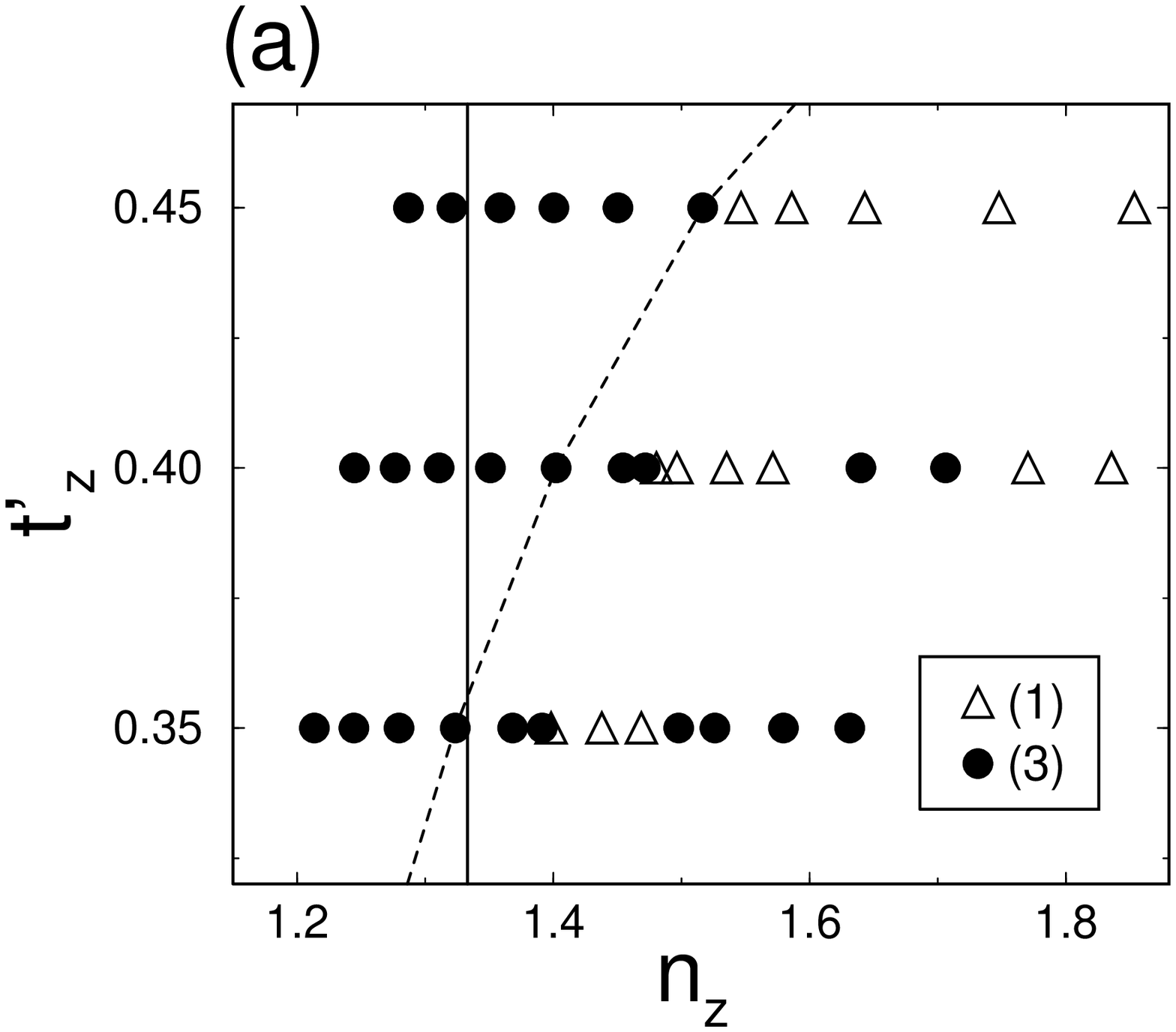} 
    \hspace{8mm}
    \includegraphics[height=6.2cm]{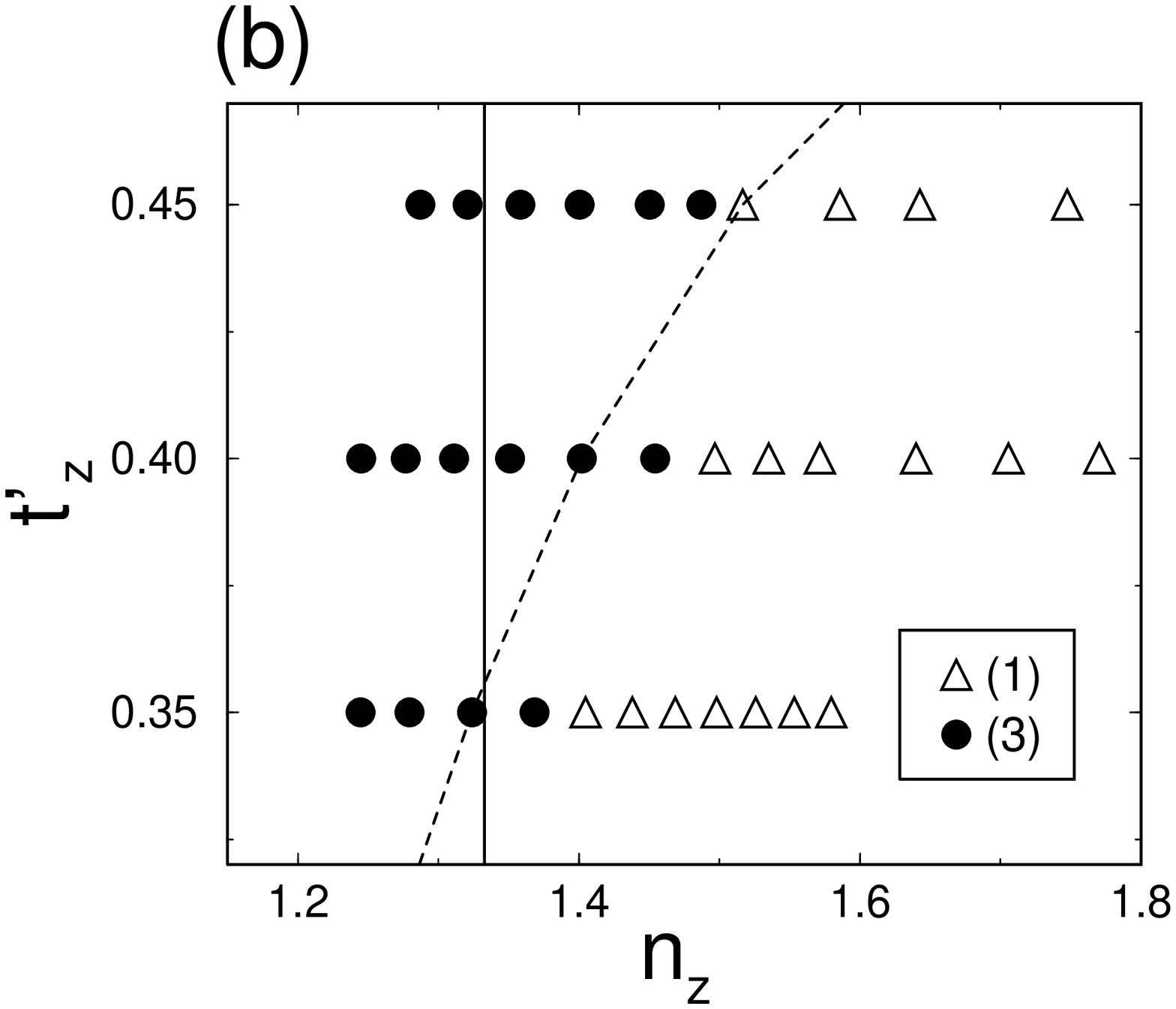} 
    \caption{Phase diagram for the parameters of the $\gamma$-band. 
             Circles and triangles represent the state 
             (3) and (1), respectively. The state (2) is not stabilized.       
             The solid line shows the number of particles $n_{{\rm z}}=1.33$. 
             The Fermi surface is 
             electron-like (hole-like) in the left (right) side 
             of the dashed line. 
             The parameters for the $\alpha'$- and $\beta'$-band are chosen as 
             (a) $t_{{\rm xy}}=1.25$, $t'_{{\rm xy}}=0.1$ and 
             (b) $t_{{\rm xy}}=1.5$, $t'_{{\rm xy}}=0.2$. 
             }
  \end{center}
\end{figure}

\section{Determination of the Pairing Symmetry}

 In this section, we compare several pairing states from the view point 
of the internal degree of freedom. 
 From the microscopic calculation in \S3 we find that 
the symmetry breaking interaction and the main attractive 
interaction have a different origin. 
 The superconductivity is mainly caused by a part of the pairing interaction 
$z_{a}(\k) V_{a}(k,k') z_{a}(\k')$, 
but this term does not contribute to the splitting of the degeneracy. 
 Instead, the $d$-vector is almost determined by the combination of 
the symmetry breaking interaction and the superconducting 
wave function $\phi_{\rm x,y}(k)$ at $\lambda=0$.  
 This argument will be valid when the perturbation with respect to 
$\lambda$ is justified. 
 In this section, we assume the pairing state at $\lambda=0$, and determine 
the $d$-vector by calculating the symmetry breaking interaction 
using the formulation in \S3.

 Before explaining the details, we summarize the obtained results in Table I 
where the stabilized state is shown for each pairing symmetry. 
 We can see that the chiral state is stabilized {\it only when} the 
symmetry is the $p$-wave and the main band is $\gamma$. 
 The other paring states give an inconsistent $d$-vector with the experimental 
result. 
 Therefore, it is concluded that the most favorable pairing state is the 
$p$-wave on the $\gamma$-band, 
which is microscopically derived in the perturbation theory (\S2).

\begin{table}[htbp]
  \begin{center}
    \caption{Obtained $d$-vector for each pairing symmetry and main band.}  
  \end{center}
\end{table}

\subsection{$P$-wave symmetry on the $\gamma$-band}

 First, we consider the pairing state obtained in \S2 and \S3. 
 We assume that the total pairing interaction is given as 
 \begin{eqnarray}
   \tilde{V}_{n,\gamma}(k,k')=\tilde{V}_{n,\gamma}^{(2)}(k,k') 
   + \tilde{V}_{\gamma}^{({\rm p})}(k,k'), 
 \end{eqnarray}
where $\tilde{V}_{n,\gamma}^{(2)}(k,k')$ is the effective interaction 
derived from the microscopic Hamiltonian within the second order. 
 We have neglected the third order terms 
($\tilde{V}_{n,\gamma}^{(3)}(k,k') \propto U^{3}$) 
because they do not contribute to the symmetry breaking interaction. 
 $\tilde{V}_{\gamma}^{({\rm p})}(k,k') = 
z_{{\rm z}}(\k)V^{({\rm p})}(k,k')z_{{\rm z}}(\k')$ 
is an attractive interaction in the $p$-wave channel, where 
we choose $ V^{({\rm p})}(k,k')=
g (\cos(k_{{\rm x}}-k'_{{\rm x}})+ \cos(k_{{\rm y}}-k'_{{\rm y}}))$. 
 If $g \leq -1$, the $p$-wave superconductivity is mainly caused by this term, 
and then we obtain the simple wave function 
$\phi_{{\rm x}}(k) \propto \sin k_{{\rm x}}$. 
 The symmetry breaking interaction is only from 
$\tilde{V}_{n,\gamma}^{(2)}(k,k')$ and determines the $d$-vector.

\begin{figure}[htbp]
  \begin{center}
    \includegraphics[height=6.5cm]{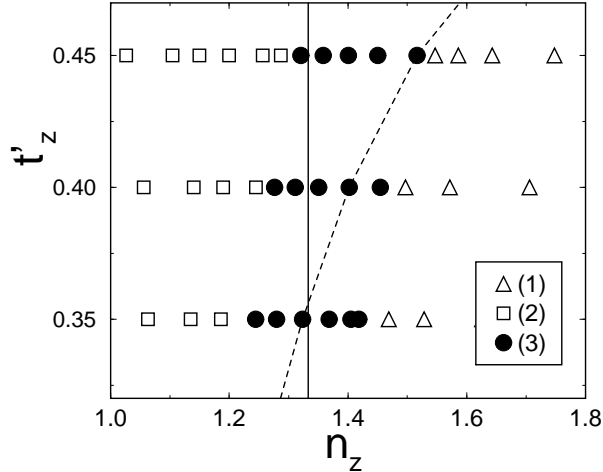} 
    \caption{Phase diagram when the $p$-wave attractive interaction 
             is assumed as eq.(27). Here, we choose $g=-3$. 
             Triangles, squares and circles represent the state 
             (1), (2) and (3), respectively. 
             The solid and dashed lines are the same as in Fig. 11. 
             }
  \end{center}
\end{figure}

 The phase diagram is shown in Fig. 12 which is qualitatively the same 
as Fig. 11. 
 Thus, the phenomenological model (eq.(27)) well reproduces the detailed 
calculation in \S3.  
 In particular, the chiral state (3) is stabilized around 
$n_{{\rm z}} \sim 1.33$. 
 The only difference is that the state (2) appears in the region 
$n_{{\rm z}} < 1.3$ where the $T_{{\rm c}}$ is very small in the perturbation 
theory. 
 We would like to stress that the chiral state (3) is robustly obtained 
independent of the detailed structure of the wave function 
$\phi_{{\rm x}}(k)$.

\subsection{$P$-wave symmetry on the $\alpha$- or $\beta$-band}

 Second, we discuss the $d$-vector when the symmetry is the $p$-wave 
but the main band is $\alpha$ or $\beta$. 
 This pairing state is also obtained by the perturbation theory when the 
DOS of $\alpha$- and $\beta$-band is large (\S2.3). 
 The notation $\alpha'$- and $\beta'$-band is convenient 
since the wave function $\phi_{{\rm x}}(k)$ has large amplitude 
on one of them. 
 When the symmetry is fixed to $d_{n}(k) = \phi_{{\rm y}}(k)$ 
($d_{n}(k) = \phi_{{\rm x}}(k)$), the main band is $\alpha'$ ($\beta'$). 
 Since the $\alpha'$-band is related to the $\beta'$-band 
through the $\pi/2$-rotation, 
$\hat{d}(k) = k_{{\rm y}}\hat{x}$ on the $\alpha'$-band is degenerate to 
$\hat{d}(k) = k_{{\rm x}}\hat{y}$ on the $\beta'$-band within 
our approximation. 
 Therefore, the same classification (1), (2) and (3) is valid.

 Since this pairing state is microscopically obtained, we show the results 
of the perturbation theory which has been formulated in \S3. 
 We similarly calculate the diagonal part of the interaction 
$\tilde{V}_{n, \beta'}(k,k')$ (or $\tilde{V}_{n, \alpha'}(k,k')$). 
 Then, the wave function on the $\beta'$-band 
has a similar momentum dependence with Fig. 2(b).

 We find that the state (2) 
$\hat{d}(k) = k_{{\rm x}} \hat{y} \pm k_{{\rm y}} \hat{x}$ is stabilized 
under the reasonable parameter set, which is inconsistent with the $\mu$SR 
experiment. 
 The splitting between the states (2) and (3) is smaller by an order of 
magnitude because the term with coefficient $U \Jh \lambda^{2}$ vanishes 
in the symmetry breaking interaction in this case. 
 Instead, the splitting energy has a coefficient 
$U' \Jh \lambda^{2}$ or $\Jh^{2} \lambda^{2}$. 
 Therefore, it is conceivable that the stabilized state changes depending 
on the details neglected in our approximation. 
 However, the weak hybridization between the $x$- and $y$-orbital generally 
stabilizes the state (1) or (2) as will be discussed in \S5. 
 It is therefore expected that the chiral state (3) is not favored 
in this case.

 These results are robust in the wide parameter region, 
$1.0 \leq n_{{\rm z}} \leq 1.6$, 
$1.26 \leq n_{{\rm x}}=n_{{\rm y}} \leq 1.64$, 
and $0.05 \leq t'_{{\rm xy}} \leq 0.5$. 
 Even if the DOS of the $\alpha'$- and $\beta'$-band is very large 
($t_{{\rm xy}}=0.7$ and $t'_{{\rm xy}}=0.056$), the stabilized $d$-vector 
does not change. 
 The qualitatively same results are obtained when we assume the attractive 
interaction as in eq.(27).

 We comment on another possibility of the $p$-wave superconductivity where 
the main band is $\alpha'$ ($\beta'$) for the wave function 
$d_{n}(k) = \phi_{{\rm x}}(k)$ 
($d_{n}(k) = \phi_{{\rm y}}(k)$). 
 In this case, nodal quasi-particles exist in all of the states (1)-(3), 
and power-law 
behaviors~\cite{rf:nishizaki,rf:bonalde,rf:ishidanode,rf:tanatar1,rf:lupien} 
can be explained. 
 We similarly investigate this possibility by assuming the pairing interaction 
like eq.(27) and conclude 
that the $d$-vector is not parallel to $\hat{z}$ 
under the reasonable parameter set (see Table. I). 
 Therefore, this candidate ($p$-wave symmetry with node) is also inconsistent 
with the time reversal symmetry breaking.~\cite{rf:luke}

\subsection{$F$-wave symmetry}

 Finally, we investigate the case of the $f$-wave symmetry which has 
gap-less quasi-particles in the chiral state. 
 For this candidate, we assume again a phenomenological pairing interaction 
like eq.(27) 
and choose 
$V^{({\rm p})}(k,k')=g (\phi_{{\rm x}}(k) \phi_{{\rm x}}(k') + 
\phi_{{\rm y}}(k) \phi_{{\rm y}}(k'))$. 
 The $f_{{\rm x}^{2}-{\rm y}^{2}}$-wave state 
is obtained by choosing   
$\phi_{{\rm x,y}}(k)=\sin k_{{\rm x,y}}(\cos k_{{\rm x}} -\cos k_{{\rm y}})$. 
 The eigenstates of the \eli equation are similarly classified into 
(1') $\hat{d}(k) = (k_{{\rm x}}^{2}-k_{{\rm y}}^{2})
(k_{{\rm x}} \hat{x} \pm k_{{\rm y}} \hat{y})$, 
(2') $\hat{d}(k) = (k_{{\rm x}}^{2}-k_{{\rm y}}^{2}) 
(k_{{\rm x}} \hat{y} \pm k_{{\rm y}} \hat{x})$ and  
(3') $\hat{d}(k) = (k_{{\rm x}}^{2}-k_{{\rm y}}^{2})
(k_{{\rm x}} \pm {\rm i} k_{{\rm y}}) \hat{z}$.

 First, we consider the case where the main band is $\gamma$. 
 In this case, the nodal quasi-particles~\cite{rf:nishizaki,rf:bonalde,
rf:ishidanode,rf:tanatar1,rf:lupien} and the time-reversal symmetry 
breaking~\cite{rf:luke} coexists in the chiral state (3'). 
 From the comparison with the thermal conductivity data, it has been proposed 
that the state (3') is the best candidate for 
${\rm Sr}_{2}{\rm Ru}{\rm O}_{4}$.~\cite{rf:dahm} 
 However, our result shows that the realized state is (2') for 
$n_{{\rm z}} >1.26$ and (1') for $n_{{\rm z}} <1.26$. 
 The state (3') is not stabilized by the spin-orbit interaction.  
 Thus, this candidate is not probable at least in the weak coupling region.

 The other candidate (3'') $\hat{d}(k) = k_{{\rm x}} k_{{\rm y}}  
(k_{{\rm x}} \pm {\rm i} k_{{\rm y}}) \hat{z}$ has been proposed by Graf and 
Balatsky.~\cite{rf:graf} 
 However, the stabilized state is not (3'') but (1'') 
$\hat{d}(k) = k_{{\rm x}} k_{{\rm y}} 
(k_{{\rm x}} \hat{x} \pm k_{{\rm y}} \hat{y})$ in our results. 
 Thus, the chiral state is not stabilized also in the $f_{{\rm xy}}$-wave 
state.

 Next, we discuss the case where the main band is $\alpha$ 
or $\beta$. 
 The candidate is $d_{n}(k) = 
\sin k_{{\rm y}} (\cos k_{{\rm x}} -\cos k_{{\rm y}})$ on the $\alpha'$-band 
(or equivalently, $d_{n}(k) = 
\sin k_{{\rm x}} (\cos k_{{\rm x}} -\cos k_{{\rm y}})$ on the $\beta'$-band). 
 We have found that this pairing state is most favorable in the triplet 
channel when $t_{{\rm xy}}/t_{{\rm z}}$ and $U/W$ is sufficiently small, 
or when the RPA or FLEX approximation is applied for small 
$t_{{\rm xy}}/t_{{\rm z}}$ (\S2). 
 We have confirmed that the stabilized state is not (3'), but  
(1') $\hat{d}(k) = (k_{{\rm x}}^{2}-k_{{\rm y}}^{2}) 
(k_{{\rm x}} \hat{x} \pm k_{{\rm y}} \hat{y})$. 
 Therefore, this candidate is also incompatible with 
the $\mu$SR experiment. 
 In Takimoto's proposal based on the orbital fluctuations, 
the main band is $\alpha'$ or $\beta'$, and the wave function is 
approximately written as $\Delta_{\alpha'}(k) = 
\sin k_{{\rm y}} (A + B (\cos k_{{\rm x}} +\cos k_{{\rm y}}))$ 
($A=0.51$ and $B=1.3$).~\cite{rf:takimoto} 
 We have also carried out the calculation for this pairing state 
and confirmed that the state 
$\hat{d}(k) = (A + B (\cos k_{{\rm x}} +\cos k_{{\rm y}})) 
(\sin k_{{\rm x}} \hat{x} \pm \sin k_{{\rm y}} \hat{y})$ is stabilized.

 The above results have been summarized in Table I. 
 It should be understood that the chiral state is stabilized only 
in the particular case.

\section{Discussions on the approximations}

 We have used the following approximations in order to determine the 
$d$-vector.

(i) Perturbation on the electron correlation.

(ii) Perturbation on the spin-orbit interaction.

(iii) Contribution from the minor band has been neglected.
 
(iv) Hybridization between the $x$-and $y$-orbitals has been neglected. 
\\
 In this section, we show that the above approximations are 
justified under the respective conditions which are expected 
for ${\rm Sr}_{2}{\rm RuO}_{4}$.

 First, the approximation (i) is a main assumption of this paper; 
the moderately weak interaction is assumed. 
 Conventional Fermi liquid behaviors~\cite{rf:maenoFermi2D,rf:mackenzie} 
have supported the weak coupling theory. 
 It is necessary for the argument in \S4 that the perturbative treatment 
for the ``symmetry breaking interaction'' is justified. 
 This assumption is more promising because the symmetry breaking interaction 
requires the Hund coupling term, and is expected to be small.

 Second, the approximation (ii) is safely justified since the spin-orbit 
interaction is sufficiently weak in ${\rm Sr}_{2}{\rm RuO}_{4}$ 
($2 \lambda \sim 0.1 {\rm eV}$) compared with  
the band width ($\sim 2 {\rm eV}$).  
 This approximation is probably not justified in the heavy fermion compounds.

 Third, we discuss the contribution from the minor band 
which is negligible in case of the ODS.~\cite{rf:agterberg} 
 In the starting model at $\lambda=0$, the order parameter in the different 
band only weakly couples through the inter-band pairing interaction, 
$V_{a,b}(k,k')$. 
 Therefore, the order parameter in the minor band has much smaller amplitude 
than that in the main band. 
 This situation is similar to the two-band superconductivity in the $s$-wave 
case.~\cite{rf:kondo} 
 The ratio furthermore decreases when the intra-band pairing interaction 
in the minor band is repulsive. 
 The contribution from the minor band to the symmetry breaking interaction has 
{\it at most} an order 
$|V_{a,b}|\lambda^{2}/W^{2} \propto J \Jh \lambda^{2}/W^{3}$, 
which is smaller than the main contribution ($\sim U \Jh \lambda^{2}/W^{3}$).  
 Thus, the approximation based on the ODS is justified 
at finite $\lambda$.

 Finally, we discuss the weak hybridization between the $x$- and $y$-orbitals. 
 Taking the hybridization into account, the unperturbed Hamiltonian eq.(17) 
is changed as, 
\begin{eqnarray}
     H'_{0} = \sum_{\k,s} 
\left(
\begin{array}{ccc}
c_{\mbox{{\scriptsize \boldmath$k$}},1,s}^{\dag} &
c_{\mbox{{\scriptsize \boldmath$k$}},2,s}^{\dag} &
c_{\mbox{{\scriptsize \boldmath$k$}},3,-s}^{\dag}\\
\end{array}
\right)
\left(
\begin{array}{ccc}
\varepsilon_{1}(\k) & {\rm i}g(\k) -s \lambda & -s \lambda\\
-{\rm i}g(\k) -s \lambda & \varepsilon_{2}(\k) &  \lambda\\
-s \lambda &  \lambda & \varepsilon_{3}(\k)\\
\end{array}
\right)
\left(
\begin{array}{c}
c_{\mbox{{\scriptsize \boldmath$k$}},1,s}\\
c_{\mbox{{\scriptsize \boldmath$k$}},2,s}\\
c_{\mbox{{\scriptsize \boldmath$k$}},3,-s}\\
\end{array}
\right). 
\end{eqnarray}
 The tight binding model gives 
$g(\k)=4 t_{{\rm m}} \sin k_{{\rm x}} \sin k_{{\rm y}}$, where  $t_{{\rm m}}$
is the hopping integral between $4d_{{\rm xz}}$- and $4d_{{\rm yz}}$-orbitals 
in the next nearest-neighbor site.  
 We can treat this mixing term in a perturbative way as is used 
for the spin-orbit interaction. 
 That is, the perturbation with respect to 
the off-diagonal term in eq.(28) is possible. 
 Two kinds of the effective interactions newly appear 
within the second order perturbation, namely the terms proportional to 
(a) $t_{{\rm m}}^{2}$ and (b) ${\rm i} t_{{\rm m}} \lambda$. 
 The term (a) is not important since it does not violate the SU(2) symmetry. 
 We discuss the cross term (b) $\propto {\rm i} t_{{\rm m}} \lambda $ 
which contributes to the symmetry breaking interaction.

 We easily find that the term (b) vanishes when the main band is $\gamma$, and 
therefore, the approximation (iv) is well justified.  
 On the contrary, the term (b) appears when the main band is $\alpha'$ or 
$\beta'$. 
 The pairing state with different reflection symmetry 
(for example, $\hat{d}(k) = k_{\rm x}\hat{x}$ on the $\alpha'$-band and 
$\hat{d}(k) = k_{\rm y}\hat{y}$ on the $\beta'$-band) 
couples through the term (b). 
 Therefore, the term (b) mainly works through the inter-band interaction 
$\tilde{V}_{\alpha',\beta'}^{{\rm (b)}}(k,k',s_{1},s_{2},s_{3},s_{4})$ 
in the present case. 
 This interaction has a finite value only when $s_{1}=s_{2}=s_{3}=s_{4}$, 
and satisfies the relation 
$\tilde{V}_{\alpha',\beta'}^{{\rm (b)}}(k,k',s)
=\tilde{V}_{\alpha',\beta'}^{{\rm (b)}}(k,k',s,s,s,s) 
=-\tilde{V}_{\alpha',\beta'}^{{\rm (b)}}(k,k'-s)$ and 
$\tilde{V}_{\alpha',\beta'}^{{\rm (b)}}(k_{{\rm x}}, k_{{\rm y}}, k'_{{\rm x}}, k'_{{\rm y}},s) = 
- \tilde{V}_{\alpha',\beta'}^{{\rm (b)}}(-k_{{\rm x}}, k_{{\rm y}}, -k'_{{\rm x}}, k'_{{\rm y}},s) = 
- \tilde{V}_{\alpha',\beta'}^{{\rm (b)}}(k_{{\rm x}}, -k_{{\rm y}}, k'_{{\rm x}}, -k'_{{\rm y}},s)
$.
Consequently, we find that the two-fold degeneracy in the eigenstates 
(1) and (2) are lifted by the term (b).  
 The other effect is negligible. 
 Then, we obtain the eigenstates 
(1a) $\hat{d}(k) = k_{{\rm x}} \hat{x} + k_{{\rm y}} \hat{y}$, 
(1b) $\hat{d}(k) = k_{{\rm x}} \hat{x} - k_{{\rm y}} \hat{y}$, 
(2a) $\hat{d}(k) = k_{{\rm x}} \hat{y} + k_{{\rm y}} \hat{x}$ and  
(2b) $\hat{d}(k) = k_{{\rm x}} \hat{y} - k_{{\rm y}} \hat{x}$ 
without the degeneracy. 
 These eigenstates correspond to the one-dimensional 
representations in the symmetry argument~\cite{rf:sureview}. 
 On the other hand, the chiral state (3) is not affected 
by the term (b), and therefore two-fold degeneracy is maintained. 
 We find that one of the states between (1a) and (1b) 
((2a) and (2b)) has a lower energy owing to the term (b), 
and the other has a higher energy. 
 This is because the matrix element arising from the term (b) has an 
opposite sign for them. 
 Consequently, two of the states (1a), (1b), (2a) and (2b) are favored 
by the term (b), and accordingly the state (3) is unfavored. 
 In other words, the $d$-vector along the $\hat{z}$-axis is not stabilized  
by the term (b). 
 Since the term (b) is in the first order with respect to $\lambda$, 
the state (3) is not stabilized in the weak coupling limit about $\lambda$ 
when the main band is $\alpha$ or $\beta$.

 As a result, the approximation (iv) does not affect the obtained 
results in Table I. 
 The mixing term works only when the main band is $\alpha$ or $\beta$. 
 Then, the state (3) is not stabilized in any case.

\section{Summary and Discussion}

 In this paper, we have theoretically investigated the internal degree 
of freedom of the triplet superconductivity in 
${\rm Sr}_{2}{\rm Ru}{\rm O}_{4}$. 
 The obtained results are summarized in the following way. 

 (1) The perturbation theory gives a $p$-wave superconductivity 
in the moderately weak coupling region. 
 The ODS is a robust result when the inter-orbital 
interaction is smaller than the intra-orbital one. 
 Then, the order parameter has an exceedingly large amplitude in one of 
the band $\alpha'$, $\beta'$ or $\gamma$. 
 The main band is $\gamma$ under the reasonable parameter set.

 (2) The spin-orbit interaction in ${\rm Ru}$ ions makes the $d$-vector 
parallel to the $\hat{z}$-axis. 
 In this case, the chiral state 
$\hat{d}(k) = (k_{{\rm x}} \pm {\rm i} k_{{\rm y}}) \hat{z}$ 
with time-reversal symmetry breaking is expected. 
 The orbital and momentum dependence of the order parameter is essential 
for this result.

 (3) When the other pairing state is assumed, 
the $d$-vector is not pointed to the $\hat{z}$-direction. 
 The $f$-wave pairing state 
and the pairing state with main $\alpha'$- or $\beta'$-band are 
included in such cases.  
 These cases are not consistent with the time reversal symmetry breaking 
indicated by the $\mu$SR experiment.~\cite{rf:luke}

 (4) The splitting of $T_{{\rm c}}$ is small 
even if the spin-orbit interaction is much larger than the characteristic 
energy of the superconductivity, namely $\lambda \gg T_{{\rm c}}, \Delta$. 
 If we put the parameters as $2\lambda =0.1{\rm eV}$ and $W =2{\rm eV}$, 
the splitting is about $\Delta T_{{\rm c}} \sim 0.04 T_{{\rm c}}$. 
 This is mainly because the ``symmetry breaking interaction'' is 
in the second order with respect to $\lambda/W$. 
 The first order term vanishes because the band mixing does not exist 
at $\lambda=0$. 
 It is an another reason that the symmetry breaking interaction 
requires the Hund coupling term. 
 Since $|\Jh|$ is smaller than the intra-band repulsion $U$, 
the symmetry breaking interaction has a smaller coefficient 
than the main part of the effective interaction.

 From the results (1)-(3), the present work has supported the weak coupling  
theory for Sr$_2$RuO$_4$.~\cite{rf:nomura,rf:nomuramulti} 
 It has been confirmed that the perturbation theory gives the probable 
pairing state, namely the $p$-wave symmetry and the main $\gamma$-band, 
and furthermore the consistent $d$-vector along the $\hat{z}$-axis.

 Then, the detailed analysis of the gap-less 
behaviors~\cite{rf:nishizaki,rf:ishidanode,rf:bonalde,rf:lupien,rf:tanatar1}
will be important for the coherent understanding. 
 If the discussion is restricted to the single band model, the line node 
appears only accidentally. 
 This situation is not improved even if we consider 
the three-dimensional model~\cite{rf:hasegawa,rf:won,rf:zhitomirski} 
because the line node is not necessarily horizontal. 
 Only the point node is allowed from the symmetry argument.~\cite{rf:bloom}
 We think that this difficulty can be resolved by the multi-band effect 
within the two-dimensional model. 
 Some proposals have been given along this line.~\cite{rf:kuroki,
rf:nomuratherm}  
 We consider that the low-energy excitation is mainly included in the 
$\alpha$- and $\beta$-band, while $T_{\rm c}$ and $d$-vector are determined by 
the $\gamma$-band. 
 To show the gap-less behaviors, an detailed investigation on the minor bands 
is essential. 
 Such study is beyond this paper and will be an important future issue.

 Here, we would like to point out that our systematic treatment 
for the spin-orbit interaction is necessary to discuss the $d$-vector. 
 The spin-orbit interaction determines the $d$-vector through 
(i) the virtual process in the effective interaction and 
(ii) the unitary transformation of the quasi-particles. 
 When the chiral state is stabilized, 
the dominant contribution comes from the cross-term of (i) and (ii). 
 Therefore, we have to treat the two effects (i) and (ii) in a same footing. 
 The estimation for the effect (i) only~\cite{rf:ogata} 
or the effect (ii) only~\cite{rf:ngls} is inadequate 
for the microscopic theory. 
 We should further mention that the effective interaction derived here is 
quite different from the phenomenological assumption in Ref. 33. 
 Indeed, the chiral state is not stabilized if we take into account 
only the effect (ii).

 At last, we comment on the phase diagram of ${\rm UPt}_{3}$ from the result 
(4). 
 A heavy Fermion system ${\rm UPt}_{3}$ is a spin triplet superconductor, and 
shows three different superconducting phases under the magnetic field.  
 The phase diagram has been explained by assuming the weak 
spin-orbit interaction~\cite{rf:ohmi} because this assumption is necessary 
to explain the NMR Knight shift.~\cite{rf:tou} 
 However, this assumption has raised a serious question since the spin-orbit 
coupling is generally strong in the heavy Fermion system.~\cite{rf:sauls}  
 We think that the result (4) gives a clue to this question.  
 It has been shown that the Hund coupling term is needed to violate 
the SU(2) symmetry in the $d$-vector space. 
 On the contrary, the SU(2) symmetry in the real spin space is violated 
even if $\Jh=0$ (see Fig. 8). 
 Therefore, the violation of the SU(2) symmetry can be much smaller 
in the $d$-vector space than in the real spin space.~\cite{rf:yanaseHF} 
 In other words, it is possible that the anisotropy is almost absorbed in the 
character of the quasi-particles, and only a weak anisotropy is remained 
in the residual interaction. 
 Thus, the microscopic theory in this paper gives a general insight on 
the triplet superconductivity.

\section*{Acknowledgments}

 The authors are grateful to Professors K. Yamada, H. Kohno, 
M. Sigrist, Y. Maeno, Y. Hasegawa, as well as 
Mr. T. Nomura and Mr. T. Koretsune for fruitful discussions. 
 Numerical computation in this work was partly carried out 
at the Yukawa Institute Computer Facility. 
 The present work was partly supported by a Grant-In-Aid for Scientific 
Research from the Ministry of Education, Science, Sports and Culture, Japan.


\begin{thebibliography}{9}
\bibitem{rf:comment21}
The result of Ref.21, which is different from ours  is probably due to 
the too large values of $U$ where the estimation of the normal self-energy 
is inaccurate.
\end{thebibliography}
\end{document}